\documentclass[aps,prl,twocolumn,preprintnumbers,superscriptaddress,nofootinbib]{revtex4-2}
\usepackage{xcolor}
\usepackage{graphicx}
\usepackage{dcolumn}   
\usepackage{bm}        
\usepackage{amssymb}   

\usepackage{float}
\usepackage{amsmath}

\usepackage{array}
\usepackage{booktabs} 

\newcommand{\seq}{\begin{subequations}}
	\newcommand{\sen}{\end{subequations}}
\newcommand{\be}{\begin{eqnarray}}
	\newcommand{\ee}{\end{eqnarray}}

\newcommand{\nn}{\nonumber}

\newcommand*{\dif}{\mathop{}\!\mathrm{d}}

\newcommand{\ba}{\begin{align}}  
	\newcommand{\ea}{\end{align}}

\DeclareMathAlphabet{\mathdutchcal}{U}{dutchcal}{m}{n}

\usepackage{tikz}
\DeclareRobustCommand{\BlueRect}{%
	\raisebox{0.6ex}{%
		\tikz[scale=0.6, baseline=(current bounding box.center)]{%
			\draw[draw=none, fill=blue!35] (0,0) rectangle (0.75,0.4);
			\draw[blue, thick] (0,0.2) -- (0.75,0.2);
		}%
	}%
}

\definecolor{MyDarkBrown}{rgb}{0.4, 0.266667, 0.133333}
\definecolor{MyDarkCyan}{rgb}{0., 0.666667, 0.666667}
\definecolor{MyLightOrange}{rgb}{1., 0.65, 0.3}
\definecolor{MyLightGreen}{rgb}{0.3, 1., 0.3}
\definecolor{MyLightMagenta}{rgb}{1., 0.3, 1.}

\DeclareRobustCommand{\BrownDashedRect}{%
	\raisebox{0.6ex}{%
		\tikz[scale=0.6, baseline=(current bounding box.center)]{%
			\fill[MyDarkBrown, fill opacity=0.4] (0,0) rectangle (0.75,0.4);
			\draw[MyDarkBrown, thick, dash pattern=on 2.1pt off 1pt] (0,0.2) -- (0.75,0.2);
		}%
	}%
}

\DeclareRobustCommand{\OrangeDashedRect}{%
	\raisebox{0.6ex}{%
		\tikz[scale=0.6, baseline=(current bounding box.center)]{%
			\fill[MyLightOrange, fill opacity=0.4] (0,0) rectangle (0.75,0.4);
			\draw[MyLightOrange, thick, dash pattern=on 1.8pt off 1pt] (0,0.2) -- (0.75,0.2);
		}%
	}%
}

\DeclareRobustCommand{\GreenDashedRect}{%
	\raisebox{0.6ex}{%
		\tikz[scale=0.6, baseline=(current bounding box.center)]{%
			\fill[MyLightGreen, fill opacity=0.4] (0,0) rectangle (0.75,0.4);
			\draw[MyLightGreen, thick, dash pattern=on 3.1pt off 2.3pt] (0,0.2) -- (0.75,0.2);
		}%
	}%
}

\DeclareRobustCommand{\MagentaDashedRect}{%
	\raisebox{0.6ex}{%
		\tikz[scale=0.6, baseline=(current bounding box.center)]{%
			\fill[MyLightMagenta, fill opacity=0.4] (0,0) rectangle (0.75,0.4);
			\draw[MyLightMagenta, thick, dash pattern=on 4pt off 3.5pt] (0,0.2) -- (0.75,0.2);
		}%
	}%
}

\DeclareRobustCommand{\BlackDotDashed}{%
	\raisebox{0.6ex}{%
		\tikz[scale=0.6, baseline=(current bounding box.center)]{%
			\draw[black, thick, dash dot] (0,0.2) -- (1,0.2);
		}%
	}%
}

\DeclareRobustCommand{\CyanDashed}{%
	\raisebox{0.6ex}{%
		\tikz[scale=0.6, baseline=(current bounding box.center)]{%
			\draw[MyDarkCyan, thick, dash pattern=on 3.3pt off 1pt] (0,0.2) -- (0.75,0.2);
		}%
	}%
}

\DeclareRobustCommand{\BlackDiamond}{%
	\raisebox{0.6ex}{%
		\tikz[scale=0.05, baseline=(current bounding box.center)]{%
			\draw[black, line width=0.7pt, fill=none] 
			(1,0) -- (2,1) -- (1,2) -- (0,1) -- cycle;
		}%
	}%
}

\DeclareRobustCommand{\RedTriangle}{%
	\raisebox{0.6ex}{%
		\tikz[scale=0.05, baseline=(current bounding box.center)]{%
			\draw[red, line width=0.7pt, fill=none] (0,0) -- (2,0) -- (1,1.732) -- cycle;
		}%
	}%
}

\DeclareRobustCommand{\BlueCircle}{%
	\raisebox{0.6ex}{%
		\tikz[scale=0.085, baseline=(current bounding box.center)]{%
			\draw[blue, line width=0.55pt, fill=none] (0,0) circle (0.5cm);
		}%
	}%
}

\DeclareRobustCommand{\GreenSquare}{%
	\raisebox{0.6ex}{%
		\tikz[scale=0.08, baseline=(current bounding box.center)]{%
			\draw[green!80!black, line width=0.6pt, fill=none] (0,0) rectangle (1cm,1cm);
		}%
	}%
}

\usepackage{lmodern}
\usepackage{hyperref}
\hypersetup{colorlinks=true,linkcolor=red,anchorcolor=blue,citecolor=blue, filecolor=blue,urlcolor=red,bookmarksnumbered=true,
	pdfview=FitB
}

\begin{document}
	
	\title{Proton Gravitational Structure and Mass Decomposition on the Light Front}
	\author{Sreeraj~Nair}
	\email{sreeraj@impcas.ac.cn}
	\affiliation{Institute of Modern Physics, Chinese Academy of Sciences, Lanzhou, Gansu, 730000, China}
	\affiliation{School of Nuclear Physics, University of Chinese Academy of Sciences, Beijing, 100049, China}
	\affiliation{CAS Key Laboratory of High Precision Nuclear Spectroscopy, Institute of Modern Physics, Chinese Academy of Sciences, Lanzhou 730000, China}
	\author{Chandan Mondal}
	\email{mondal@impcas.ac.cn} 
	\affiliation{Institute of Modern Physics, Chinese Academy of Sciences, Lanzhou, Gansu, 730000, China}
	\affiliation{School of Nuclear Physics, University of Chinese Academy of Sciences, Beijing, 100049, China}
	\affiliation{CAS Key Laboratory of High Precision Nuclear Spectroscopy, Institute of Modern Physics, Chinese Academy of Sciences, Lanzhou 730000, China}
	\author{Siqi~Xu}
	\email{xsq234@impcas.ac.cn}
	\affiliation{Institute of Modern Physics, Chinese Academy of Sciences, Lanzhou, Gansu, 730000, China}
	\affiliation{School of Nuclear Physics, University of Chinese Academy of Sciences, Beijing, 100049, China}
	\affiliation{CAS Key Laboratory of High Precision Nuclear Spectroscopy, Institute of Modern Physics, Chinese Academy of Sciences, Lanzhou 730000, China}
	\author{Xingbo Zhao}
	\email{xbzhao@impcas.ac.cn} 
	\affiliation{Institute of Modern Physics, Chinese Academy of Sciences, Lanzhou, Gansu, 730000, China}
	\affiliation{School of Nuclear Physics, University of Chinese Academy of Sciences, Beijing, 100049, China}
	\affiliation{CAS Key Laboratory of High Precision Nuclear Spectroscopy, Institute of Modern Physics, Chinese Academy of Sciences, Lanzhou 730000, China}
	\affiliation{Advanced Energy Science and Technology Guangdong Laboratory, Huizhou, Guangdong 516000, China}
	\author{James P. Vary}
	\email{jvary@iastate.edu} 
	\affiliation{Department of Physics and Astronomy, Iowa State University, Ames, Iowa 50011, USA}
	
	\collaboration{BLFQ Collaboration}

	\begin{abstract}
		Gravitational form factors (GFFs) of hadrons encode essential information about the internal distributions of mass, spin, pressure, and shear among their quark and gluon constituents. We compute the quark and gluon GFFs of the proton using a fully relativistic, nonperturbative framework based on a light-front quantized Hamiltonian with quantum chromodynamics (QCD) input. This allows us to quantify the impact of a dynamical gluon on the proton's mechanical properties, such as pressure and shear distributions. Our predictions agree well with recent lattice QCD results and experimental extractions. We also determine the proton's mass and mechanical radii and address the long-standing puzzle of its mass decomposition. At the scale $\mu^2 = 4~\mathrm{GeV}^2$, we find that quark energy, gluon field energy, the quark condensate, and the QCD trace anomaly contribute $31.5\%$, $34.7\%$, $11.3\%$, and $22.5\%$, respectively, which are consistent with lattice QCD findings.
		
	\end{abstract}
	
	\maketitle
	
	{\it Introduction}.---Understanding the internal structure of the proton, a bound state of quarks and gluons governed by quantum chromodynamics (QCD), remains a central challenge in nuclear and particle physics~\cite{AbdulKhalek:2021gbh,Anderle:2021wcy,Accardi:2023chb}. Gravitational form factors (GFFs), defined via the proton matrix elements of the QCD energy-momentum tensor (EMT), provide access to fundamental mechanical properties such as the internal distributions of mass, spin, pressure, and shear~\cite{Lorce:2018egm,Polyakov:2002yz,Polyakov:2018zvc,Burkert:2023wzr}.
	
	The GFFs are related to the second Mellin moments of generalized parton distributions (GPDs), and can be constrained through data from deeply virtual Compton scattering and meson production~\cite{Diehl:2003ny,Ji:1996ek,dHose:2016mda}. Quark GFFs have been extracted from Jefferson Lab measurements~\cite{Burkert:2018bqq}, with future constraints expected from PANDA at FAIR~\cite{PANDA:2009yku}, EICs~\cite{AbdulKhalek:2021gbh,Anderle:2021wcy}, NICA~\cite{MPD:2022qhn}, ILC, and J-PARC~\cite{Kumano:2022cje}. Theoretically, quark GFFs have been studied using lattice QCD, chiral perturbation theory, light-cone sum rules, QCD-inspired models, and light-front quantization~\cite{Chen:2001pva,Hagler:2003jd,Pasquini:2014vua,Abidin:2009hr,Nair:2024fit}.
	
	In contrast, the gluon GFFs remain poorly constrained, both experimentally and theoretically. Near-threshold charmonium and bottomonium production offers a promising probe of the gluonic structure of the proton~\cite{Mamo:2019mka,Guo:2023qgu,Hatta:2018ina,GlueX:2023pev}, with recent $J/\psi$ data from JLab and prospects at future EICs~\cite{Duran:2022xag,Liu:2024yqa,Hechenberger:2024abg}. These processes are sensitive to the proton's gluon GPDs through two-gluon exchange, enabling indirect access to gluon GFFs under factorization~\cite{Guo:2021ibg,Hatta:2021can}. Analyses based on GPD fits and lattice QCD suggest dipole- or tripole-like behaviors for these form factors~\cite{Pefkou:2021fni,Hackett:2023rif}.
	
	While many theoretical models focus only on quark degrees of freedom, the role of gluons is essential for a complete picture of the proton's mechanical structure. In particular, the gluon contributions to the $D$-term, relevant for pressure and shear forces, depend on quark-gluon interactions that are encoded in the ``bad" components of the EMT~\cite{Hu:2024edc}. Existing theoretical work on gluon GFFs includes lattice QCD~\cite{Shanahan:2018nnv,Alexandrou:2020sml}, holographic models~\cite{Mamo:2021krl}, and recent Bethe-Salpeter equation approaches~\cite{Yao:2024ixu}. Improved theoretical determinations of these quantities are essential for refining future measurements and advancing our understanding of the proton's structure.
	
	In this work, we address the proton's gravitational structure using basis light-front quantization (BLFQ), a fully relativistic and nonperturbative framework for solving quantum field theory problems~\cite{Vary:2009gt,Mondal:2019jdg,Xu:2021wwj,Xu:2022yxb}. We employ an effective light-front Hamiltonian incorporating QCD interactions~\cite{Brodsky:1997de} with explicit quark ($q$) and gluon ($g$) degrees of freedom in the $|qqq\rangle$ and $|qqqg\rangle$ Fock sectors, along with a complementary three-dimensional confinement potential~\cite{Li:2015zda}. By solving for the mass eigenstates at low-resolution scales, we obtain the light-front wave functions (LFWFs) and extract the quark and gluon GFFs of the proton.

	This approach allows us to address two key aspects of proton structure. First, it enables a quantitative assessment of the impact of a dynamical gluon on the proton's mechanical properties, including the pressure and shear force distributions. Second, it provides a nonperturbative determination of the proton's mass decomposition into contributions from quark energy, gluon field energy, the quark scalar condensate, and the QCD trace anomaly, which can only be reliably extracted through QCD-based calculations and experimental input.
	
	{\it Proton wave functions from light-front QCD Hamiltonian}.---In the light-front (LF) formalism, the proton state is expanded in terms of Fock sectors as  
	$
	|\Psi\rangle = \psi_{(qqq)}|qqq\rangle + \psi_{(qqqg)}|qqqg\rangle + \cdots,
	$
	where the LFWFs $\psi_{(\cdots)}$ encode the probability amplitudes for different partonic configurations. These LFWFs are obtained by solving the Hamiltonian eigenvalue equation,  
	$
	(P^+P^- - \vec{P}^{\perp2})|\Psi\rangle = M^2|\Psi\rangle,
	$  
	where $P^\pm=P^0 \pm P^3$ define the longitudinal momentum ($P^+$) and the LF Hamiltonian ($P^-$) with $M^2$ the mass squared eigenvalue.
	
	At the initial scale, where the proton consists of $|qqq\rangle$ and $|qqqg\rangle$ components, we use the LF Hamiltonian $P^- = P^-_{\rm QCD} + P^-_{\rm I}$, with $P^-_{\rm QCD}$ encoding the relevant QCD interactions and $P^-_{\rm I}$ modeling confinement~\cite{Xu:2022yxb}. 
	In the LF gauge ($A^+=0$), with one dynamical gluon~\cite{Xu:2023nqv,Lan:2021wok},  
	\begin{align}
		P_{\rm QCD}^- =&\int dx^- d^2 x^\perp\Big\{\frac{1}{2}\bar{\psi}\gamma^+\frac{m_0^2+(i\partial^\perp)^2}{i\partial^+}\psi \nonumber\\
		&+\frac{1}{2}A^i_a\left[m_g^2+(i\partial^\perp)^2\right]A^i_a + g_c\bar{\psi}\gamma_\mu T^a A^\mu_a\psi \nonumber\\
		&+\frac{g_c^2}{2}\bar{\psi}\gamma^+ T^a\psi\frac{1}{(i\partial^+)^2}\bar{\psi}\gamma^+ T^a\psi\Big\},
	\end{align}
	where $\psi$ and $A_a^i$ are quark and gluon fields, with bare masses $m_0$, $m_g$, and coupling $g_c$. $T$ denotes the generators of the $SU(3)$ color gauge group, and $\gamma^\mu$ are the Dirac matrices. A phenomenological gluon mass $m_g$ accounts for non-perturbative effects~\cite{PhysRevD.26.1453}. Quark masses are renormalized via $m_q=m_0-\delta m_q$~\cite{Perry:1990mz,Karmanov:2008br}, with additional non-perturbative contributions parameterized by  effective vertex mass $m_f$~\cite{Burkardt:1998dd}.
	
	Confinement is implemented in the leading Fock sector~\cite{Li:2015zda}
	\begin{equation}
		P^-_{\rm I}P^+ = \frac{\kappa^4}{2}\sum_{i\neq j}\left[\vec{r}_{ij}^{\perp2}-\frac{\partial_{x_i}(x_i x_j\partial_{x_j})}{(m_i+m_j)^2}\right],
	\end{equation}
	with transverse separation $\vec{r}_{ij}^{\perp2}$ and strength $\kappa$. We omit explicit confinement in the $|qqqg\rangle$ sector, anticipating that the restricted transverse basis and inclusion of a massive gluon effectively capture essential confinement features at this stage.

	We solve the LF Hamiltonian eigenvalue problem using the BLFQ framework~\cite{Vary:2009gt}, expanding the proton state in a basis composed of plane waves in the longitudinal direction (within a box of length $2L$ with antiperiodic/periodic boundary conditions for quarks/gluons), two-dimensional harmonic oscillator (2D-HO) functions $\Phi_{nm}(\vec{p}^\perp; b)$ in the transverse plane~\cite{Zhao:2014xaa}, and light-cone helicity spinors. Each single-particle basis state is labeled by $\bar{\alpha} = \{k, n, m, \lambda\}$, where $k$ is the longitudinal mode (half-integer for quarks, integer for gluons, excluding zero mode), $n$ and $m$ are 2D-HO quantum numbers, and $\lambda$ is helicity. Fock sectors with multiple color-singlet combinations, such as $|qqqg\rangle$, require an additional label.
	
	The basis is truncated using $N_{\rm max}$ and $\mathcal{K}$ to render the Hamiltonian matrix finite. $N_{\rm max}$ limits the total 2D-HO energy: $\sum_i (2n_i + |m_i| + 1) \leq N_{\rm max}$, while $\mathcal{K} = \sum_i k_i$ sets the longitudinal resolution, with momentum fractions given by $x_i = k_i/\mathcal{K}$.
	
	Numerical truncations $\mathcal{K}$ and $\mathcal{N}_{\mathrm{max}}$ define infrared and ultraviolet scales~\cite{Zhao:2014xaa}. Diagonalizing the Hamiltonian gives the proton's momentum-space wavefunctions with helicity $\Lambda$:
	\begin{equation}
		\Psi^{N,\Lambda}_{\{x_i,\vec{p}_{\perp i},\lambda_i\}}=\sum_{\{n_i m_i\}}\psi^{N}(\{\alpha_i\})\prod_{i=1}^{N}\Phi_{n_i m_i}(\vec{p}_{i\perp},b),
	\end{equation}
	where coefficients $\psi^{N=3}$ and $\psi^{N=4}$ are the components of the eigenvectors associated with $|uud\rangle$ and $|uudg\rangle$.
	
	\begin{figure*}[htp!]
		\centering
		\includegraphics[width=18cm,height=7cm,clip]{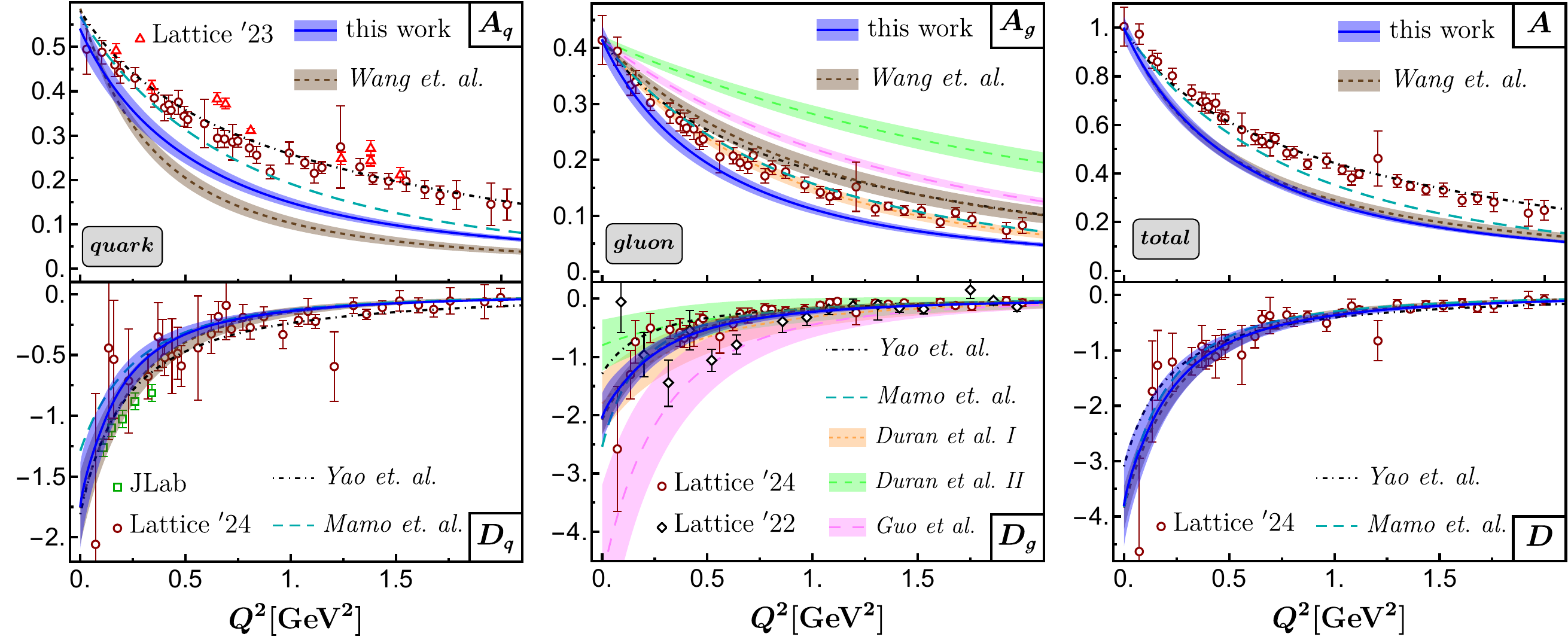}
		\caption{Proton GFFs $A(Q^2)$ and $D(Q^2)$ at $\mu^2=4~\mathrm{GeV}^2$, separated into quark ($\mathdutchcal{q}$) and gluon ($\mathdutchcal{g}$) contributions. Our results (\protect\BlueRect) are compared with lattice QCD (\protect\BlackDiamond~\cite{Pefkou:2021fni}, \protect\RedTriangle~\cite{Bhattacharya:2023ays}, \protect\BlueCircle~\cite{Hackett:2023rif}), rescaled JLab data for $D_\mathdutchcal{q}(Q^2)$ (\protect\GreenSquare~\cite{Burkert:2018bqq,Shanahan:2018nnv}), and theoretical models: vector meson photoproduction (\protect\BrownDashedRect~\cite{Wang:2023fmx}), Faddeev equation (\protect\BlackDotDashed~\cite{Yao:2024ixu}), string-based model (\protect\CyanDashed~\cite{Mamo:2024jwp}), and multipole parametrizations (\protect\OrangeDashedRect~\protect\GreenDashedRect~\cite{Duran:2022xag}, \protect \MagentaDashedRect~\cite{Guo:2023pqw}). All $A_{\mathdutchcal{g}}(Q^2)$ curves--ours and all comparisons--are rescaled to $A_{\mathdutchcal{g}}(0)=0.414$. The shaded band on our results indicates a $10\%$ uncertainty originating from the initial scale.
		}
		\label{fig1}
	\end{figure*}
	
	{\it Gravitational form factors}.---The GFFs parameterize the EMT $\theta^{\mu\nu}$ for a spin-$1/2$ composite system as~\cite{Polyakov:2018zvc}:
	\begin{align}
		&\langle P^{\prime},\Lambda^{\prime}|\, \theta^{\mu\nu}_i(0)\,|P,\Lambda \rangle = \ \overline{u}(P^{\prime},\Lambda^{\prime}) \Big[ A_i(Q^2) \frac{\overline{P}^{\mu}\ \overline{P}^{\nu}}{M} \nonumber \\
		&+ J_i(Q^2) \frac{i \overline{P}^{\{\mu} \sigma^{\nu\} \alpha}q_\alpha}{2M} + D_i(Q^2) \frac{q^{\mu} q^{\nu} - q^2 g^{\mu\nu}}{4M} \nonumber \\
		&+ \overline{C}_i(Q^2) M g^{\mu\nu} \Big] u(P,\Lambda),
	\end{align}
	where $a^{\{ \mu}b^{\nu \}} = (a^\mu b^\nu + a^\nu b^\mu)/2$. Here, $\overline{P}^{\mu}=\frac{1}{2}(P^{\prime}+P)^{\mu}$; $\overline{u}(P',\Lambda^{\prime}),u(P,\Lambda)$ are Dirac spinors, $M$ is the proton mass, and $Q^2 = -q^2$ is the squared momentum transfer in the transverse direction. Indices $i=(\mathdutchcal{q},\mathdutchcal{g})$ distinguish quark and gluon contributions, respectively, and helicities $(\Lambda,\Lambda^{\prime})=(\uparrow,\downarrow)$ indicate spin orientations along the $z$-axis. The form factor $J_i(Q^2)$ relates to $A_i$ and $B_i$ by $J_i=(A_i+B_i)/2$.
	
	The symmetric QCD EMT decomposes into quark and gluon parts, $\theta^{\mu \nu} = \sum_{i=(\mathdutchcal{q},\mathdutchcal{g})}\theta^{\mu \nu}_i$, with 
	$\theta^{\mu \nu}_\mathdutchcal{q} = i\overline{\psi}\gamma^{\{\mu}\mathcal{D}^{\nu\}}\psi$ and $\theta^{\mu \nu}_\mathdutchcal{g} = - F^{\mu \lambda a}F_{\lambda a}^{\nu} + \frac{1}{4} g^{\mu \nu} ( F_{\lambda \sigma a})^2$, where $\mathcal{D}^\mu =\partial^\mu +ig A^\nu$ is the covariant derivative.
	
	The matrix elements of the EMT, required to extract the GFFs, can be expressed compactly as $	\mathcal{M}^{\mu \nu }_{\Lambda \Lambda^{\prime}} =\langle P',\Lambda^{\prime}|\,\theta^{\mu \nu }_i(0)\,|P,\Lambda \rangle$. The GFFs $ A_i(Q^2)$ and $B_i(Q^2)$ are then obtained from the longitudinal components as follows:
	\begin{equation}
		\begin{aligned}\label{rhsA}
			\mathcal{M}^{++}_{\uparrow \uparrow} + \mathcal{M}^{++}_{\downarrow \downarrow} &= 2\,(P^+)^2 A_i(Q^2)\,, \\
			\mathcal{M}^{++}_{\uparrow \downarrow} + \mathcal{M}^{++}_{\downarrow \uparrow} &= \frac{i q^{(2)}}{M}(P^+)^2 B_i(Q^2)\,.
		\end{aligned}
	\end{equation}
	The transverse components yield the GFFs $D_i(Q^2)$ and $\overline{C}_i(Q^2)$:
	\begin{equation}
		\begin{aligned}\label{rhsC}
			\mathcal{M}^{12}_{\uparrow \uparrow} + \mathcal{M}^{12}_{\downarrow \downarrow} &= D_i(Q^2) q^{(1)} q^{(2)}, \\ 
			q_{\mu}\mathcal{M}^{\mu 1 }_{\uparrow \downarrow } + q_{\mu}\mathcal{M}^{\mu 1 }_{ \downarrow \uparrow}&= 2 i M q^{(1)} q^{(2)} \overline{C}_i(Q^2).
		\end{aligned}
	\end{equation}
	The GFF \(\overline{C}_i(Q^2)\) in Eq.~\eqref{rhsC} arises from the non-conservation of the partial EMT~\cite{Lorce:2018egm}. These four GFFs can be expressed explicitly in terms of overlaps of the LFWFs as:

	\begin{equation}
		\begin{aligned}\label{eq:olpa}
			A_i(Q^2) &= \frac{1}{2}\int_N x_1 
			{\bf{\Psi}}^{\prime\Lambda*}{\bf{\Psi}}^{\Lambda}, \\
			iq^{(2)} B_i(Q^2) &= M \int_N x_1 
			{\bf{\Psi}}^{\prime\Lambda*}{\bf{\Psi}}^{-\Lambda}, \\
			q^{(1)} q^{(2)} C_i(Q^2) &= \int_N  \frac{\mathcal{O}^{12}_{i,\lambda \lambda'}}{x_1}
			{\bf{\Psi}}^{\prime\Lambda*}{\bf{\Psi}}^{\Lambda}, \\
			2 i M q^{(1)} q^{(2)} \overline{C}_i(Q^2) &=  \int_N   \frac{q^{(k)} \mathcal{O}^{1k}_{i,\lambda \lambda'}}{x_1}
			{\bf{\Psi}}^{\prime\Lambda*}{\bf{\Psi}}^{-\Lambda},
		\end{aligned}
	\end{equation}
	where the struck parton's momenta are $x'_1 = x_1$, $\vec{p}{\,'}_{i}^{\perp}= \vec p_1^\perp + (1 - x_1)\,\vec q^\perp$, while for spectators $x'_i = x_i$, $\vec{p}{\,'}_{i}^{\perp}= {\vec{p}}_{i}^{\perp} - x_i \vec{q}^\perp$. The initial and final LFWFs are denoted $\boldsymbol{\Psi}^\Lambda = \Psi^{N,\Lambda}_{\{x_i,\vec{p}_{i}^{\perp},\lambda_i\}}$ and $\boldsymbol{\Psi}^{\prime\Lambda*} = \Psi^{N,\Lambda*}_{\{x'_i,\vec{p}{\,'}_{i}^{\perp},\lambda'_i\}}$ and the form of $\mathcal{O}_{i,\lambda \lambda'}$ is expressed as:
	\begin{align}
		\mathcal{O}^{12 }_{\mathdutchcal{q},\lambda \lambda'} &= 
		\delta_{\lambda \lambda'} 	\Big[
		2\,\mathcal{R}^{(1)}\,\mathcal{R}^{(2)}
		+ i\,\lambda \left(\mathcal{R}^{(1)}\,q^{(1)}-\mathcal{R}^{(2)}\,q^{(2)}\right)
		\Big] , \nonumber\\
		\mathcal{O}^{11 }_{\mathdutchcal{q},\lambda \lambda'} &= 
		\delta_{\lambda\lambda'}\,
		\mathcal{R}^{(1)}\left(\mathcal{R}^{(1)}-i\,\lambda\,q^{(2)}\right) , \nonumber\\
		\mathcal{O}^{12 }_{\mathdutchcal{g},\lambda \lambda'} &= 
		\tfrac12\, \delta_{\lambda\lambda'}
		\Big[ \mathcal{R}^{(1)}\mathcal{R}^{(2)}
		+ \mathcal{R}^{(1)}q^{(2)}+q^{(1)}\mathcal{R}^{(2)}
		+q^{(1)}q^{(2)}	\Big] \nonumber \\ &+  \tfrac12\,i\lambda
		\Big[
		\left(\mathcal{R}^{(2)}+q^{(2)}\right)q^{(2)}-\left(\mathcal{R}^{(1)}+q^{(1)}\right)q^{(1)}
		\Big], \nonumber\\
		\mathcal{O}^{11}_{\mathdutchcal{g},\lambda \lambda'} &= 
		\tfrac12\, \delta_{\lambda\lambda'}
		\left(\mathcal{R}^{(1)} + q^{(1)}\right)^2
		+ i\lambda \left(p^{(2)}q^{(1)}-p^{(1)}q^{(2)}\right)
		\nn \\ & 	-   i\lambda \left( \mathcal{R}^{(1)}q^{(2)} + \mathcal{R}^{(2)} q^{(1)}\right)\delta_{\lambda\lambda'}	 \label{eq:op} 
	\end{align}
	where $	\mathcal{R}^{(j)} \equiv 2p^{(j)} + (1-x_1) q^{(j)}$. The integration measure in Eq.~\eqref{eq:olpa} is expressed as:
	$	\int_N \equiv \sum_{\substack{N,\Lambda, \lambda_i,\lambda'_i}} \prod_{i=1}^N \frac{[\dif x_i\, \dif^2 p_{i}^{\perp}]}{(16\pi^3)^{N-1}} \delta\!\left(1\!-\!\sum_j x_j\right)\delta^{(2)}\!\left(\sum_j \vec{p}_{j}^{\perp}\right).\nn
	$

	All calculations are performed with basis truncations $\mathcal{N}_{\rm max}=9$, $\mathcal{K}=16.5$, HO scale $b=0.70~\mathrm{GeV}$, and UV cutoff for the instantaneous interaction $b_{\mathrm{inst}}=3.00~\mathrm{GeV}$. The Hamiltonian parameters $\{m_u,m_d,m_g,\kappa,m_f,g_c\}=\{0.31,0.25,0.50,0.54,1.80,2.40\}$ (in GeV, except $g_c$) reproduce the proton mass and its electromagnetic properties~\cite{Xu:2023nqv}. We find quark probabilities of $\sim 44\%$ in $|qqq\rangle$ and $\sim 56\%$ in $|qqqg\rangle$ at the model scale.
	
	Figure~\ref{fig1} presents the proton GFFs $A(Q^2)$ and $D(Q^2)$ at $\mu^2 = 4\,\mathrm{GeV}^2$, showing separate quark and gluon contributions. We compare our results with lattice QCD and experimental extractions based on various phenomenological approaches. Our $D(Q^2)$ and its quark and gluon components agree well with lattice QCD~\cite{Hackett:2023rif}, while $A(Q^2)$, particularly the quark contribution, shows some deviation at high $Q^2$~\cite{Hackett:2023rif,Bhattacharya:2023ays}. The quark form factor $D_{\mathdutchcal{q}}(Q^2)$ agrees with JLab data~\cite{Shanahan:2018nnv,Burkert:2018bqq}, and the gluon form factor $D_{\mathdutchcal{g}}(Q^2)$ aligns with extractions from near-threshold vector meson photoproduction~\cite{Duran:2022xag,Guo:2023pqw,GlueX:2023pev,Wang:2023fmx}, as well as with string-based predictions~\cite{Mamo:2024jwp} and Faddeev-equation results~\cite{Yao:2024ixu}. Our gluon GFFs are consistent with experimentally extracted results based on the holographic QCD approach (Duran et al. I~\cite{Duran:2022xag}), but differ from those obtained using a GPD-inspired analysis (Duran et al. II~\cite{Duran:2022xag}). A recent GPD fit to GlueX data~\cite{Guo:2023pqw,GlueX:2023pev} shows better agreement with our $A_{\mathdutchcal{g}}(Q^2)$. Since $B(Q^2) \approx 0$, we have $J(Q^2) \approx A(Q^2)/2$ (see Supplemental Material). At $Q^2 = 0$, we find:
	$A_{\mathdutchcal{q}}(0) = 0.54(04)$, $A_{\mathdutchcal{g}}(0) = 0.46(03)$, and $D_{\mathdutchcal{q}}(0) = -1.73(35)$, $D_{\mathdutchcal{g}}(0) = -2.04(41)$. Consequently, the total $D$-term at $Q^2=0$ is $D(0) = -3.77 \pm 0.74$. A detailed comparison of our total $D(0)$ with lattice QCD and other theoretical studies is presented in the Supplemental Material.

	\begin{figure}[htp!]
		\centering
		\includegraphics[width=8.0cm,height=5.5cm,clip]{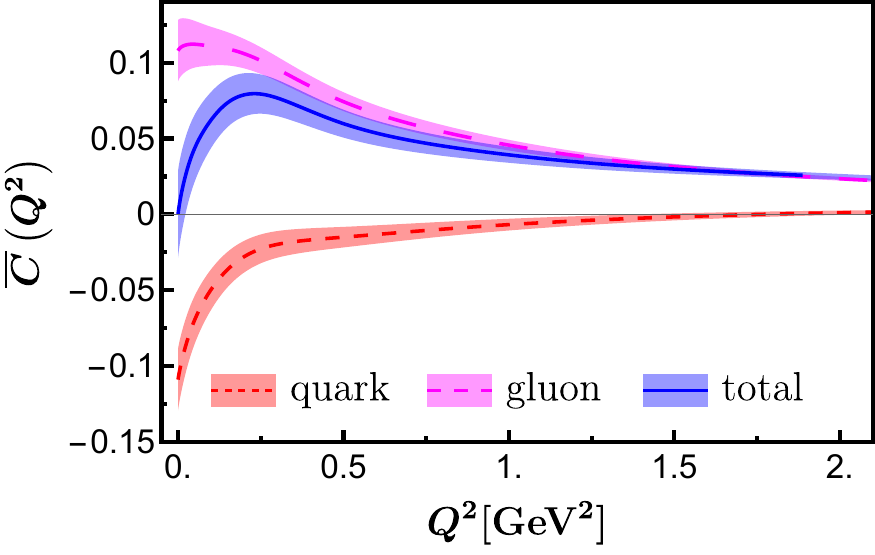}
		\caption{Proton's GFF $\overline{C}(Q^2)$ and its quark and gluon
			components as functions of $Q^2$.} 
		\label{fig2} 
	\end{figure}

	Figure~\ref{fig2} presents GFF $\overline{C}(Q^2)$, highlighting opposite signs for quark ($\overline{C}_\mathdutchcal{q}<0$) and gluon ($\overline{C}_\mathdutchcal{g}>0$). The sum rule $\overline{C}(Q^2)=0$ is approximately satisfied at $Q^2=0$: $\overline{C}(0)\approx 0$, with $\overline{C}_\mathdutchcal{q}(0)=-0.11(02)$, $\overline{C}_\mathdutchcal{g}(0)=0.11(02)$.

	{\it Mechanical densities and radii}.---Defined through the EMT and by analogy with classical mechanical systems, the GFF $D(Q^2)$ provides insight into the internal pressure and shear force distributions in the proton~\cite{Polyakov:2018zvc}. These mechanical distributions are expressed as~\cite{Kim:2021jjf}:
	\begin{equation}
		\begin{aligned}
			\label{Prefun}
			\mathcal{P}(b) &=\frac{1}{16M}\frac{1}{b}\frac{d}{db} \left[b \frac{d}{db} \tilde{D}(b)\right], \\
			\mathcal{S}(b) &=-\frac{1}{8M} b \frac{d}{db}\left[  \frac{1}{b} \frac{d}{db} \tilde{D}(b)\right],
		\end{aligned}
	\end{equation}
	where $ \tilde{D}(b) = \frac{1}{(2\pi)^2}\,\int d^2 \vec{q}^{\perp} \ e^{-i \vec{q}^{\perp} \cdot\vec{b}^{\perp} } D(q^2)$ and $b=|{\vec{b}^{\perp}}|$.
	
	\begin{figure}[htp!]
		\centering
		\includegraphics[width=7cm,height=7.5cm,clip]{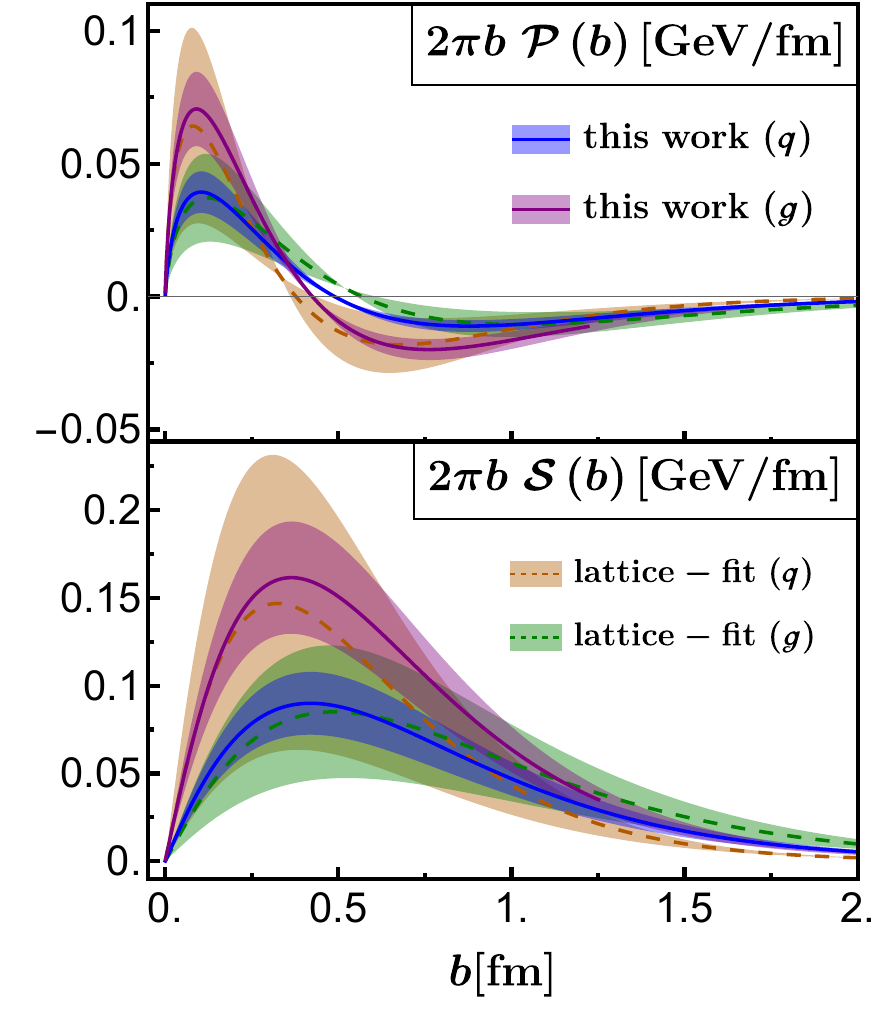}
		\caption{Pressure (upper) and shear (lower) distributions compared with lattice QCD results. The ``lattice--fit" curves are obtained using the dipole fit parameters from Ref.~\cite{Hackett:2023rif}.}\label{fig3} 
	\end{figure}
	
	Figure~\ref{fig3} shows the pressure ($2\pi b~ \mathcal{P}(b)$) and shear ($2\pi b~ \mathcal{S}(b)$) distributions, derived using the fitted GFF parameters provided in the Supplemental Material. The pressure distribution is positive at small $b$ and negative at larger $b$, consistent with mechanical equilibrium~\cite{Polyakov:2018zvc}, while the shear distribution remains positive, indicating local stability~\cite{Polyakov:2018zvc}. For comparison, we use lattice dipole fit parameters from Ref.~\cite{Hackett:2023rif} to perform the two-dimensional Fourier transform, with results labeled ``lattice--fit" in Fig.~\ref{fig3}. Our quark distributions have lower peak magnitudes compared to the lattice fit, whereas our gluon distributions exhibit higher peaks. Overall, though magnitudes differ, the qualitative behavior agrees with the lattice-based results.
	
\begin{figure}[htp!]
		\centering
		\includegraphics[width=8.4cm,height=7.0cm,clip]{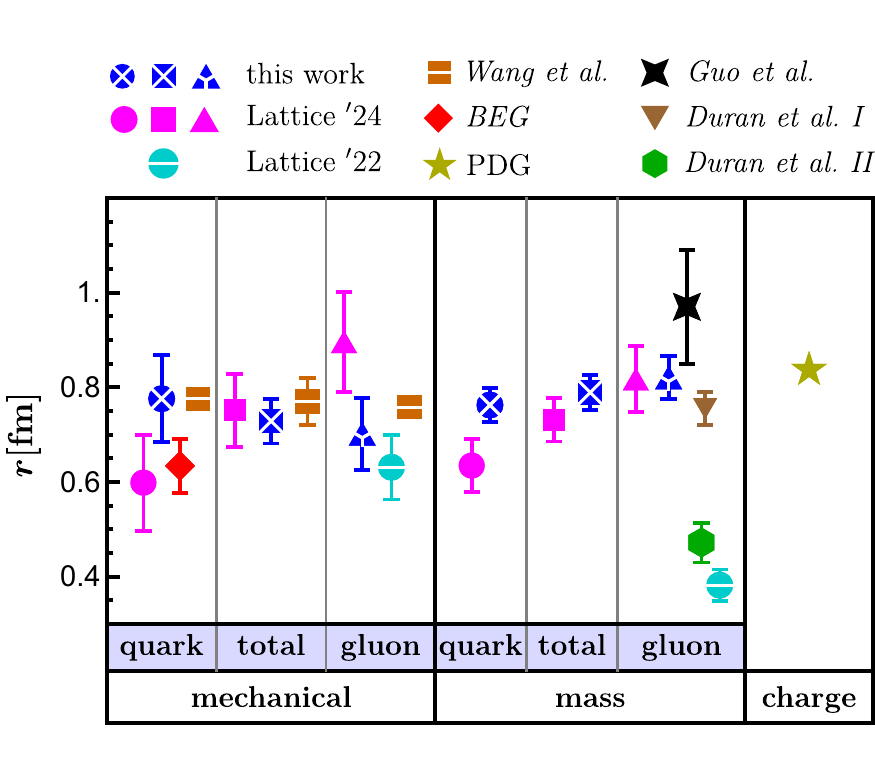}
		\caption{Proton mechanical and mass radii compared with lattice QCD results~\cite{Pefkou:2021fni,Hackett:2023rif}. The quark mechanical radius is shown alongside the BEG result~\cite{Burkert:2018bqq}, the gluon mass radius with extractions by Duran et al.~\cite{Duran:2022xag} and Guo et al.~\cite{Guo:2023pqw}, and the total mechanical radius with Wang et al.~\cite{Wang:2023fmx}. The charge radius is from PDG~\cite{ParticleDataGroup:2024cfk}.}\label{fig4} 
	\end{figure}

	The mechanical and mass radii are determined from the GFFs~\cite{Polyakov:2018zvc,Duran:2022xag}:
	\begin{equation}
		\begin{aligned}\label{radius}
			\langle r^2_{\text{mech}}\rangle &=6 D(0) \Big[ \int^{\infty}_{0} {\rm d}Q^2~ D(Q^2)\Big]^{-1}, \\
			\langle r^2_{\text{mass}}\rangle &= \frac{6}{A(0)}\frac{dA}{dt}\Big{|}_{t=0} - \frac{3}{2A(0)}\frac{D(0)}{M^2}.
		\end{aligned}
	\end{equation}
	We obtain quark mechanical and mass radii of $0.78(9)$ fm and $0.76(4)$ fm, respectively, consistent with lattice and experimental results~\cite{Hackett:2023rif,Burkert:2018bqq}. The gluon mechanical radius, $0.70(8)$ fm, lies between existing lattice estimates~\cite{Hackett:2023rif,Pefkou:2021fni}, while the gluon mass radius, $0.82(5)$ fm, agrees with recent lattice and phenomenological studies~\cite{Hackett:2023rif,Duran:2022xag,Guo:2023pqw}. The total mechanical and mass radii, $0.73(5)$ fm and $0.79(4)$ fm, are in close agreement with lattice results~\cite{Hackett:2023rif} and vector meson photoproduction data~\cite{Wang:2023fmx}, and are slightly below the proton charge radius~\cite{ParticleDataGroup:2024cfk}. See Fig.~\ref{fig4} for comparison.

	{\it Proton mass decompositions}.---Ji's mass decomposition~\cite{Ji:1994av} separates proton mass ($M$) into quark energy ($M_q$), gluon field energy ($M_g$), quark mass ($M_m$), and the QCD trace anomaly ($M_a$) contributions:
	$M = M_q + M_g + M_m + M_a$,
	with explicit terms
	$M_q = \frac{3}{4}(a - b/(1+\gamma_m))M$; $M_g = \frac{3}{4}(1 - a)M$; $M_m = \frac{1}{4}\frac{4 + \gamma_m}{1 + \gamma_m}\,b\,M$; and $M_a = \frac{1}{4}(1 - b)\,M$, where the parameters $a = A_\mathdutchcal{q}(0)$, $b = A_\mathdutchcal{q}(0) + 4\overline{C}_\mathdutchcal{q}(0)$.  
	Alternatively, the proton mass can be expressed in terms of internal energies ($U_i = [A_i(0) + \overline{C}_i(0)]M$) as $M = U_q + U_g$~\cite{Lorce:2017xzd}.

\begin{figure}[htp!]
		\centering
		\includegraphics[width=4.2cm,height=2.8cm,clip]{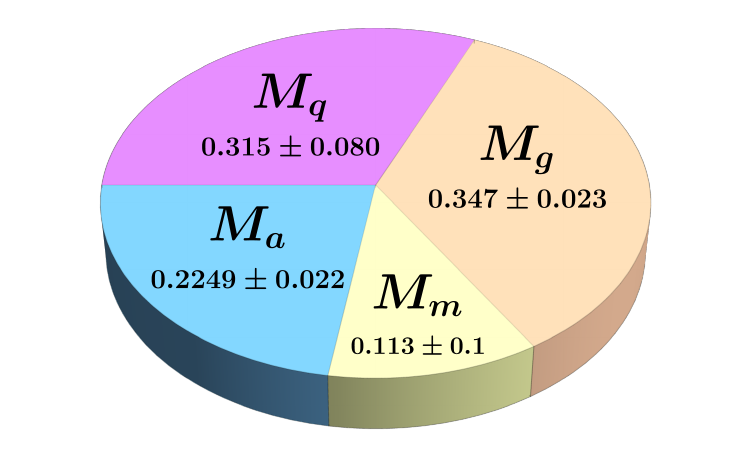}
		\includegraphics[width=4.2cm,height=2.8cm,clip]{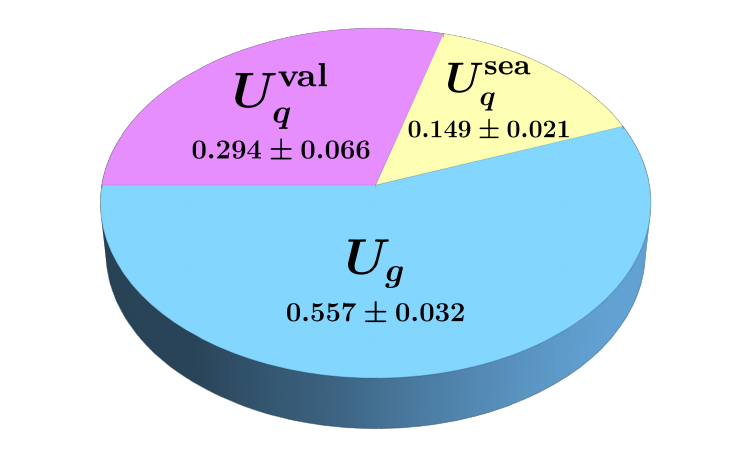}
		\caption{Proton mass decomposition following Ji's four-term scheme \cite{Ji:1994av} and the two-term scheme of Ref.~\cite{Lorce:2017xzd}. Values are given in units of $M$ at $\mu^2 = 4~\mathrm{GeV}^2$. In the two-term decomposition, the valence and sea quark contributions are shown separately, with $U_q = U_q^{\mathrm{val}} + U_q^{\mathrm{sea}}$.}
		\label{fig5} 
	\end{figure}

	Figure~\ref{fig5} shows the proton mass decompositions, highlighting significant gluon dominance. In Ji's four-term decomposition, the gluon, quark energy, quark mass, and trace anomaly contributions are $34.7\%$, $31.5\%$, $11.3\%$, and $22.5\%$, respectively. We adopt an anomalous quark mass dimension of $\gamma_m\approx -0.15$ for $n_f = 3$ active flavors~\cite{Lorce:2017xzd}.  The two-term decomposition yields a gluon contribution of $55.7\%$, with quark contributions split into valence ($29.4\%$) and sea ($14.9\%$) components. Table~\ref{table:mass_decomp_compare} compares our four-term results at $\mu^2 = 4~\mathrm{GeV}^2$ with lattice QCD findings~\cite{Yang:2018nqn}, showing good agreement. In both cases, the gluon contribution is dominant, followed by quark energy, trace anomaly, and quark mass terms.
	
	\begin{table}[htp!]
		\centering
		\setlength{\tabcolsep}{6pt}
		\caption{Proton mass decomposition using Ji's four-term scheme~\cite{Ji:1994av} at $\mu^2 = 4\,\mathrm{GeV}^2$, presented as percentages of the proton mass $M$. Our results are compared with those from lattice QCD~\cite{Yang:2018nqn}.}
		\label{table:mass_decomp_compare}
		\begin{ruledtabular}
			\begin{tabular}{lcccc}
				& $M_q~(\%)$ & $M_g~(\%)$ & $M_m~(\%)$ & $M_a~(\%)$ \\ \hline
				This work & 31.5 & 34.7 & 11.3 & 22.5 \\
				Lattice~\cite{Yang:2018nqn} & 32 & 36 & 9 & 23 \\
			\end{tabular}
		\end{ruledtabular}
	\end{table}

	{\it Summary}.---We solved the light-front QCD Hamiltonian for the proton, including both the constituent three-quark and the three-quark--one-gluon Fock sectors. The resulting light-front wave functions were used to compute the quark and gluon contributions to the proton's GFFs  and mass decomposition. We quantified the role of a dynamical gluon in shaping the proton's mechanical properties, including pressure and shear distributions. Our predictions show good agreement with recent lattice QCD results and experimental extractions across several key aspects of the proton's internal structure.
	
	We extracted the proton's mass and mechanical radii as $0.79(4)$ fm and $0.73(5)$ fm, respectively. The corresponding quark contributions are $0.76(4)$ fm and $0.78(9)$ fm, while the gluon contributions are $0.82(5)$ fm and $0.70(8)$ fm. These values align well with recent lattice and phenomenological analyses.
	
	We also provided a nonperturbative determination of the proton's mass decomposition, finding contributions from quark energy ($31.5\%$), gluon field energy ($34.7\%$), the quark condensate ($11.3\%$), and the QCD trace anomaly ($22.5\%$), consistent with lattice QCD. Additionally, we analyzed the decomposition using an internal-energy-based two-term scheme, confirming the dominant role of gluons ($55.7\%$) in the proton mass.
	
{\it Acknowledgments}.---C.M. acknowledges support from the new faculty startup funding provided by the Institute of Modern Physics, Chinese Academy of Sciences (Grant No. E129952YR0). X.Z. is supported by the new faculty startup funding at the Institute of Modern Physics, Chinese Academy of Sciences; the Key Research Program of Frontier Sciences, Chinese Academy of Sciences (Grant No. ZDB-SLY-7020); the Foundation for Key Talents of Gansu Province; the Central Funds Guiding Local Science and Technology Development of Gansu Province (Grant No. 22ZY1QA006); the National Natural Science Foundation of China (Grant No. 12375143); the National Key R\&D Program of China (Grant No. 2023YFA1606903); and the Strategic Priority Research Program of the Chinese Academy of Sciences (Grant No. XDB34000000). This research also receives support from the Gansu International Collaboration and Talents Recruitment Base of Particle Physics (2023?2027), the Senior Scientist Program funded by Gansu Province (Grant No. 25RCKA008), and the International Partnership Program of the Chinese Academy of Sciences (Grant No. 016GJHZ2022103FN). J.P.V. is supported by the U.S. Department of Energy (Grant No. DE-SC0023692). Computational resources were primarily provided by the Sugon Advanced Computing Center.

\bibliographystyle{apsrev4-2}
\bibliography{ref.bib}

\begin{thebibliography}{57}%
\makeatletter
\providecommand \@ifxundefined [1]{%
 \@ifx{#1\undefined}
}%
\providecommand \@ifnum [1]{%
 \ifnum #1\expandafter \@firstoftwo
 \else \expandafter \@secondoftwo
 \fi
}%
\providecommand \@ifx [1]{%
 \ifx #1\expandafter \@firstoftwo
 \else \expandafter \@secondoftwo
 \fi
}%
\providecommand \natexlab [1]{#1}%
\providecommand \enquote  [1]{``#1''}%
\providecommand \bibnamefont  [1]{#1}%
\providecommand \bibfnamefont [1]{#1}%
\providecommand \citenamefont [1]{#1}%
\providecommand \href@noop [0]{\@secondoftwo}%
\providecommand \href [0]{\begingroup \@sanitize@url \@href}%
\providecommand \@href[1]{\@@startlink{#1}\@@href}%
\providecommand \@@href[1]{\endgroup#1\@@endlink}%
\providecommand \@sanitize@url [0]{\catcode `\\12\catcode `\$12\catcode
  `\&12\catcode `\#12\catcode `\^12\catcode `\_12\catcode `\%12\relax}%
\providecommand \@@startlink[1]{}%
\providecommand \@@endlink[0]{}%
\providecommand \url  [0]{\begingroup\@sanitize@url \@url }%
\providecommand \@url [1]{\endgroup\@href {#1}{\urlprefix }}%
\providecommand \urlprefix  [0]{URL }%
\providecommand \Eprint [0]{\href }%
\providecommand \doibase [0]{https://doi.org/}%
\providecommand \selectlanguage [0]{\@gobble}%
\providecommand \bibinfo  [0]{\@secondoftwo}%
\providecommand \bibfield  [0]{\@secondoftwo}%
\providecommand \translation [1]{[#1]}%
\providecommand \BibitemOpen [0]{}%
\providecommand \bibitemStop [0]{}%
\providecommand \bibitemNoStop [0]{.\EOS\space}%
\providecommand \EOS [0]{\spacefactor3000\relax}%
\providecommand \BibitemShut  [1]{\csname bibitem#1\endcsname}%
\let\auto@bib@innerbib\@empty
\bibitem [{\citenamefont {Abdul~Khalek}\ \emph {et~al.}(2022)\citenamefont
  {Abdul~Khalek} \emph {et~al.}}]{AbdulKhalek:2021gbh}%
  \BibitemOpen
  \bibfield  {author} {\bibinfo {author} {\bibfnamefont {R.}~\bibnamefont
  {Abdul~Khalek}} \emph {et~al.},\ }\href
  {https://doi.org/10.1016/j.nuclphysa.2022.122447} {\bibfield  {journal}
  {\bibinfo  {journal} {Nucl. Phys. A}\ }\textbf {\bibinfo {volume} {1026}},\
  \bibinfo {pages} {122447} (\bibinfo {year} {2022})},\ \Eprint
  {https://arxiv.org/abs/2103.05419} {arXiv:2103.05419 [physics.ins-det]}
  \BibitemShut {NoStop}%
\bibitem [{\citenamefont {Anderle}\ \emph {et~al.}(2021)\citenamefont {Anderle}
  \emph {et~al.}}]{Anderle:2021wcy}%
  \BibitemOpen
  \bibfield  {author} {\bibinfo {author} {\bibfnamefont {D.~P.}\ \bibnamefont
  {Anderle}} \emph {et~al.},\ }\href
  {https://doi.org/10.1007/s11467-021-1062-0} {\bibfield  {journal} {\bibinfo
  {journal} {Front. Phys. (Beijing)}\ }\textbf {\bibinfo {volume} {16}},\
  \bibinfo {pages} {64701} (\bibinfo {year} {2021})},\ \Eprint
  {https://arxiv.org/abs/2102.09222} {arXiv:2102.09222 [nucl-ex]} \BibitemShut
  {NoStop}%
\bibitem [{\citenamefont {Accardi}\ \emph {et~al.}(2023)\citenamefont {Accardi}
  \emph {et~al.}}]{Accardi:2023chb}%
  \BibitemOpen
  \bibfield  {author} {\bibinfo {author} {\bibfnamefont {A.}~\bibnamefont
  {Accardi}} \emph {et~al.},\ }\href@noop {} {\  (\bibinfo {year} {2023})},\
  \Eprint {https://arxiv.org/abs/2306.09360} {arXiv:2306.09360 [nucl-ex]}
  \BibitemShut {NoStop}%
\bibitem [{\citenamefont {Lorc\'e}\ \emph {et~al.}(2019)\citenamefont
  {Lorc\'e}, \citenamefont {Moutarde},\ and\ \citenamefont
  {Trawi\'nski}}]{Lorce:2018egm}%
  \BibitemOpen
  \bibfield  {author} {\bibinfo {author} {\bibfnamefont {C.}~\bibnamefont
  {Lorc\'e}}, \bibinfo {author} {\bibfnamefont {H.}~\bibnamefont {Moutarde}},\
  and\ \bibinfo {author} {\bibfnamefont {A.~P.}\ \bibnamefont {Trawi\'nski}},\
  }\href {https://doi.org/10.1140/epjc/s10052-019-6572-3} {\bibfield  {journal}
  {\bibinfo  {journal} {Eur. Phys. J. C}\ }\textbf {\bibinfo {volume} {79}},\
  \bibinfo {pages} {89} (\bibinfo {year} {2019})},\ \Eprint
  {https://arxiv.org/abs/1810.09837} {arXiv:1810.09837 [hep-ph]} \BibitemShut
  {NoStop}%
\bibitem [{\citenamefont {Polyakov}(2003)}]{Polyakov:2002yz}%
  \BibitemOpen
  \bibfield  {author} {\bibinfo {author} {\bibfnamefont {M.~V.}\ \bibnamefont
  {Polyakov}},\ }\href {https://doi.org/10.1016/S0370-2693(03)00036-4}
  {\bibfield  {journal} {\bibinfo  {journal} {Phys. Lett. B}\ }\textbf
  {\bibinfo {volume} {555}},\ \bibinfo {pages} {57} (\bibinfo {year} {2003})},\
  \Eprint {https://arxiv.org/abs/hep-ph/0210165} {arXiv:hep-ph/0210165}
  \BibitemShut {NoStop}%
\bibitem [{\citenamefont {Polyakov}\ and\ \citenamefont
  {Schweitzer}(2018)}]{Polyakov:2018zvc}%
  \BibitemOpen
  \bibfield  {author} {\bibinfo {author} {\bibfnamefont {M.~V.}\ \bibnamefont
  {Polyakov}}\ and\ \bibinfo {author} {\bibfnamefont {P.}~\bibnamefont
  {Schweitzer}},\ }\href {https://doi.org/10.1142/S0217751X18300259} {\bibfield
   {journal} {\bibinfo  {journal} {Int. J. Mod. Phys. A}\ }\textbf {\bibinfo
  {volume} {33}},\ \bibinfo {pages} {1830025} (\bibinfo {year} {2018})},\
  \Eprint {https://arxiv.org/abs/1805.06596} {arXiv:1805.06596 [hep-ph]}
  \BibitemShut {NoStop}%
\bibitem [{\citenamefont {Burkert}\ \emph {et~al.}(2023)\citenamefont
  {Burkert}, \citenamefont {Elouadrhiri}, \citenamefont {Girod}, \citenamefont
  {Lorc\'e}, \citenamefont {Schweitzer},\ and\ \citenamefont
  {Shanahan}}]{Burkert:2023wzr}%
  \BibitemOpen
  \bibfield  {author} {\bibinfo {author} {\bibfnamefont {V.~D.}\ \bibnamefont
  {Burkert}}, \bibinfo {author} {\bibfnamefont {L.}~\bibnamefont
  {Elouadrhiri}}, \bibinfo {author} {\bibfnamefont {F.~X.}\ \bibnamefont
  {Girod}}, \bibinfo {author} {\bibfnamefont {C.}~\bibnamefont {Lorc\'e}},
  \bibinfo {author} {\bibfnamefont {P.}~\bibnamefont {Schweitzer}},\ and\
  \bibinfo {author} {\bibfnamefont {P.~E.}\ \bibnamefont {Shanahan}},\ }\href
  {https://doi.org/10.1103/RevModPhys.95.041002} {\bibfield  {journal}
  {\bibinfo  {journal} {Rev. Mod. Phys.}\ }\textbf {\bibinfo {volume} {95}},\
  \bibinfo {pages} {041002} (\bibinfo {year} {2023})},\ \Eprint
  {https://arxiv.org/abs/2303.08347} {arXiv:2303.08347 [hep-ph]} \BibitemShut
  {NoStop}%
\bibitem [{\citenamefont {Diehl}(2003)}]{Diehl:2003ny}%
  \BibitemOpen
  \bibfield  {author} {\bibinfo {author} {\bibfnamefont {M.}~\bibnamefont
  {Diehl}},\ }\href {https://doi.org/10.1016/j.physrep.2003.08.002} {\bibfield
  {journal} {\bibinfo  {journal} {Phys. Rept.}\ }\textbf {\bibinfo {volume}
  {388}},\ \bibinfo {pages} {41} (\bibinfo {year} {2003})},\ \Eprint
  {https://arxiv.org/abs/hep-ph/0307382} {arXiv:hep-ph/0307382} \BibitemShut
  {NoStop}%
\bibitem [{\citenamefont {Ji}(1997)}]{Ji:1996ek}%
  \BibitemOpen
  \bibfield  {author} {\bibinfo {author} {\bibfnamefont {X.-D.}\ \bibnamefont
  {Ji}},\ }\href {https://doi.org/10.1103/PhysRevLett.78.610} {\bibfield
  {journal} {\bibinfo  {journal} {Phys. Rev. Lett.}\ }\textbf {\bibinfo
  {volume} {78}},\ \bibinfo {pages} {610} (\bibinfo {year} {1997})},\ \Eprint
  {https://arxiv.org/abs/hep-ph/9603249} {arXiv:hep-ph/9603249} \BibitemShut
  {NoStop}%
\bibitem [{\citenamefont {d'Hose}\ \emph {et~al.}(2016)\citenamefont {d'Hose},
  \citenamefont {Niccolai},\ and\ \citenamefont {Rostomyan}}]{dHose:2016mda}%
  \BibitemOpen
  \bibfield  {author} {\bibinfo {author} {\bibfnamefont {N.}~\bibnamefont
  {d'Hose}}, \bibinfo {author} {\bibfnamefont {S.}~\bibnamefont {Niccolai}},\
  and\ \bibinfo {author} {\bibfnamefont {A.}~\bibnamefont {Rostomyan}},\ }\href
  {https://doi.org/10.1140/epja/i2016-16151-9} {\bibfield  {journal} {\bibinfo
  {journal} {Eur. Phys. J. A}\ }\textbf {\bibinfo {volume} {52}},\ \bibinfo
  {pages} {151} (\bibinfo {year} {2016})}\BibitemShut {NoStop}%
\bibitem [{\citenamefont {Burkert}\ \emph {et~al.}(2018)\citenamefont
  {Burkert}, \citenamefont {Elouadrhiri},\ and\ \citenamefont
  {Girod}}]{Burkert:2018bqq}%
  \BibitemOpen
  \bibfield  {author} {\bibinfo {author} {\bibfnamefont {V.~D.}\ \bibnamefont
  {Burkert}}, \bibinfo {author} {\bibfnamefont {L.}~\bibnamefont
  {Elouadrhiri}},\ and\ \bibinfo {author} {\bibfnamefont {F.~X.}\ \bibnamefont
  {Girod}},\ }\href {https://doi.org/10.1038/s41586-018-0060-z} {\bibfield
  {journal} {\bibinfo  {journal} {Nature}\ }\textbf {\bibinfo {volume} {557}},\
  \bibinfo {pages} {396} (\bibinfo {year} {2018})}\BibitemShut {NoStop}%
\bibitem [{\citenamefont {Lutz}\ \emph {et~al.}(2009)\citenamefont {Lutz} \emph
  {et~al.}}]{PANDA:2009yku}%
  \BibitemOpen
  \bibfield  {author} {\bibinfo {author} {\bibfnamefont {M.~F.~M.}\
  \bibnamefont {Lutz}} \emph {et~al.} (\bibinfo {collaboration} {PANDA}),\
  }\href@noop {} {\  (\bibinfo {year} {2009})},\ \Eprint
  {https://arxiv.org/abs/0903.3905} {arXiv:0903.3905 [hep-ex]} \BibitemShut
  {NoStop}%
\bibitem [{\citenamefont {Abgaryan}\ \emph {et~al.}(2022)\citenamefont
  {Abgaryan} \emph {et~al.}}]{MPD:2022qhn}%
  \BibitemOpen
  \bibfield  {author} {\bibinfo {author} {\bibfnamefont {V.}~\bibnamefont
  {Abgaryan}} \emph {et~al.} (\bibinfo {collaboration} {MPD}),\ }\href
  {https://doi.org/10.1140/epja/s10050-022-00750-6} {\bibfield  {journal}
  {\bibinfo  {journal} {Eur. Phys. J. A}\ }\textbf {\bibinfo {volume} {58}},\
  \bibinfo {pages} {140} (\bibinfo {year} {2022})},\ \Eprint
  {https://arxiv.org/abs/2202.08970} {arXiv:2202.08970 [physics.ins-det]}
  \BibitemShut {NoStop}%
\bibitem [{\citenamefont {Kumano}(2022)}]{Kumano:2022cje}%
  \BibitemOpen
  \bibfield  {author} {\bibinfo {author} {\bibfnamefont {S.}~\bibnamefont
  {Kumano}},\ }\href {https://doi.org/10.3390/physics4020037} {\bibfield
  {journal} {\bibinfo  {journal} {MDPI Physics}\ }\textbf {\bibinfo {volume}
  {4}},\ \bibinfo {pages} {565} (\bibinfo {year} {2022})},\ \Eprint
  {https://arxiv.org/abs/2205.03012} {arXiv:2205.03012 [hep-ph]} \BibitemShut
  {NoStop}%
\bibitem [{\citenamefont {Chen}\ and\ \citenamefont {Ji}(2002)}]{Chen:2001pva}%
  \BibitemOpen
  \bibfield  {author} {\bibinfo {author} {\bibfnamefont {J.-W.}\ \bibnamefont
  {Chen}}\ and\ \bibinfo {author} {\bibfnamefont {X.-d.}\ \bibnamefont {Ji}},\
  }\href {https://doi.org/10.1103/PhysRevLett.88.052003} {\bibfield  {journal}
  {\bibinfo  {journal} {Phys. Rev. Lett.}\ }\textbf {\bibinfo {volume} {88}},\
  \bibinfo {pages} {052003} (\bibinfo {year} {2002})},\ \Eprint
  {https://arxiv.org/abs/hep-ph/0111048} {arXiv:hep-ph/0111048} \BibitemShut
  {NoStop}%
\bibitem [{\citenamefont {Hagler}\ \emph {et~al.}(2003)\citenamefont {Hagler},
  \citenamefont {Negele}, \citenamefont {Renner}, \citenamefont {Schroers},
  \citenamefont {Lippert},\ and\ \citenamefont {Schilling}}]{Hagler:2003jd}%
  \BibitemOpen
  \bibfield  {author} {\bibinfo {author} {\bibfnamefont {P.}~\bibnamefont
  {Hagler}}, \bibinfo {author} {\bibfnamefont {J.~W.}\ \bibnamefont {Negele}},
  \bibinfo {author} {\bibfnamefont {D.~B.}\ \bibnamefont {Renner}}, \bibinfo
  {author} {\bibfnamefont {W.}~\bibnamefont {Schroers}}, \bibinfo {author}
  {\bibfnamefont {T.}~\bibnamefont {Lippert}},\ and\ \bibinfo {author}
  {\bibfnamefont {K.}~\bibnamefont {Schilling}} (\bibinfo {collaboration}
  {LHPC, SESAM}),\ }\href {https://doi.org/10.1103/PhysRevD.68.034505}
  {\bibfield  {journal} {\bibinfo  {journal} {Phys. Rev. D}\ }\textbf {\bibinfo
  {volume} {68}},\ \bibinfo {pages} {034505} (\bibinfo {year} {2003})},\
  \Eprint {https://arxiv.org/abs/hep-lat/0304018} {arXiv:hep-lat/0304018}
  \BibitemShut {NoStop}%
\bibitem [{\citenamefont {Pasquini}\ \emph {et~al.}(2014)\citenamefont
  {Pasquini}, \citenamefont {Polyakov},\ and\ \citenamefont
  {Vanderhaeghen}}]{Pasquini:2014vua}%
  \BibitemOpen
  \bibfield  {author} {\bibinfo {author} {\bibfnamefont {B.}~\bibnamefont
  {Pasquini}}, \bibinfo {author} {\bibfnamefont {M.~V.}\ \bibnamefont
  {Polyakov}},\ and\ \bibinfo {author} {\bibfnamefont {M.}~\bibnamefont
  {Vanderhaeghen}},\ }\href {https://doi.org/10.1016/j.physletb.2014.10.047}
  {\bibfield  {journal} {\bibinfo  {journal} {Phys. Lett. B}\ }\textbf
  {\bibinfo {volume} {739}},\ \bibinfo {pages} {133} (\bibinfo {year}
  {2014})},\ \Eprint {https://arxiv.org/abs/1407.5960} {arXiv:1407.5960
  [hep-ph]} \BibitemShut {NoStop}%
\bibitem [{\citenamefont {Abidin}\ and\ \citenamefont
  {Carlson}(2009)}]{Abidin:2009hr}%
  \BibitemOpen
  \bibfield  {author} {\bibinfo {author} {\bibfnamefont {Z.}~\bibnamefont
  {Abidin}}\ and\ \bibinfo {author} {\bibfnamefont {C.~E.}\ \bibnamefont
  {Carlson}},\ }\href {https://doi.org/10.1103/PhysRevD.79.115003} {\bibfield
  {journal} {\bibinfo  {journal} {Phys. Rev. D}\ }\textbf {\bibinfo {volume}
  {79}},\ \bibinfo {pages} {115003} (\bibinfo {year} {2009})},\ \Eprint
  {https://arxiv.org/abs/0903.4818} {arXiv:0903.4818 [hep-ph]} \BibitemShut
  {NoStop}%
\bibitem [{\citenamefont {Nair}\ \emph {et~al.}(2024)\citenamefont {Nair},
  \citenamefont {Mondal}, \citenamefont {Xu}, \citenamefont {Zhao},
  \citenamefont {Mukherjee},\ and\ \citenamefont {Vary}}]{Nair:2024fit}%
  \BibitemOpen
  \bibfield  {author} {\bibinfo {author} {\bibfnamefont {S.}~\bibnamefont
  {Nair}}, \bibinfo {author} {\bibfnamefont {C.}~\bibnamefont {Mondal}},
  \bibinfo {author} {\bibfnamefont {S.}~\bibnamefont {Xu}}, \bibinfo {author}
  {\bibfnamefont {X.}~\bibnamefont {Zhao}}, \bibinfo {author} {\bibfnamefont
  {A.}~\bibnamefont {Mukherjee}},\ and\ \bibinfo {author} {\bibfnamefont
  {J.~P.}\ \bibnamefont {Vary}} (\bibinfo {collaboration} {BLFQ}),\ }\href
  {https://doi.org/10.1103/PhysRevD.110.056027} {\bibfield  {journal} {\bibinfo
   {journal} {Phys. Rev. D}\ }\textbf {\bibinfo {volume} {110}},\ \bibinfo
  {pages} {056027} (\bibinfo {year} {2024})},\ \Eprint
  {https://arxiv.org/abs/2403.11702} {arXiv:2403.11702 [hep-ph]} \BibitemShut
  {NoStop}%
\bibitem [{\citenamefont {Mamo}\ and\ \citenamefont
  {Zahed}(2020)}]{Mamo:2019mka}%
  \BibitemOpen
  \bibfield  {author} {\bibinfo {author} {\bibfnamefont {K.~A.}\ \bibnamefont
  {Mamo}}\ and\ \bibinfo {author} {\bibfnamefont {I.}~\bibnamefont {Zahed}},\
  }\href {https://doi.org/10.1103/PhysRevD.101.086003} {\bibfield  {journal}
  {\bibinfo  {journal} {Phys. Rev. D}\ }\textbf {\bibinfo {volume} {101}},\
  \bibinfo {pages} {086003} (\bibinfo {year} {2020})},\ \Eprint
  {https://arxiv.org/abs/1910.04707} {arXiv:1910.04707 [hep-ph]} \BibitemShut
  {NoStop}%
\bibitem [{\citenamefont {Guo}\ \emph {et~al.}(2024)\citenamefont {Guo},
  \citenamefont {Ji},\ and\ \citenamefont {Yuan}}]{Guo:2023qgu}%
  \BibitemOpen
  \bibfield  {author} {\bibinfo {author} {\bibfnamefont {Y.}~\bibnamefont
  {Guo}}, \bibinfo {author} {\bibfnamefont {X.}~\bibnamefont {Ji}},\ and\
  \bibinfo {author} {\bibfnamefont {F.}~\bibnamefont {Yuan}},\ }\href
  {https://doi.org/10.1103/PhysRevD.109.014014} {\bibfield  {journal} {\bibinfo
   {journal} {Phys. Rev. D}\ }\textbf {\bibinfo {volume} {109}},\ \bibinfo
  {pages} {014014} (\bibinfo {year} {2024})},\ \Eprint
  {https://arxiv.org/abs/2308.13006} {arXiv:2308.13006 [hep-ph]} \BibitemShut
  {NoStop}%
\bibitem [{\citenamefont {Hatta}\ and\ \citenamefont
  {Yang}(2018)}]{Hatta:2018ina}%
  \BibitemOpen
  \bibfield  {author} {\bibinfo {author} {\bibfnamefont {Y.}~\bibnamefont
  {Hatta}}\ and\ \bibinfo {author} {\bibfnamefont {D.-L.}\ \bibnamefont
  {Yang}},\ }\href {https://doi.org/10.1103/PhysRevD.98.074003} {\bibfield
  {journal} {\bibinfo  {journal} {Phys. Rev. D}\ }\textbf {\bibinfo {volume}
  {98}},\ \bibinfo {pages} {074003} (\bibinfo {year} {2018})},\ \Eprint
  {https://arxiv.org/abs/1808.02163} {arXiv:1808.02163 [hep-ph]} \BibitemShut
  {NoStop}%
\bibitem [{\citenamefont {Adhikari}\ \emph {et~al.}(2023)\citenamefont
  {Adhikari} \emph {et~al.}}]{GlueX:2023pev}%
  \BibitemOpen
  \bibfield  {author} {\bibinfo {author} {\bibfnamefont {S.}~\bibnamefont
  {Adhikari}} \emph {et~al.} (\bibinfo {collaboration} {GlueX}),\ }\href
  {https://doi.org/10.1103/PhysRevC.108.025201} {\bibfield  {journal} {\bibinfo
   {journal} {Phys. Rev. C}\ }\textbf {\bibinfo {volume} {108}},\ \bibinfo
  {pages} {025201} (\bibinfo {year} {2023})},\ \Eprint
  {https://arxiv.org/abs/2304.03845} {arXiv:2304.03845 [nucl-ex]} \BibitemShut
  {NoStop}%
\bibitem [{\citenamefont {Duran}\ \emph {et~al.}(2023)\citenamefont {Duran}
  \emph {et~al.}}]{Duran:2022xag}%
  \BibitemOpen
  \bibfield  {author} {\bibinfo {author} {\bibfnamefont {B.}~\bibnamefont
  {Duran}} \emph {et~al.},\ }\href {https://doi.org/10.1038/s41586-023-05730-4}
  {\bibfield  {journal} {\bibinfo  {journal} {Nature}\ }\textbf {\bibinfo
  {volume} {615}},\ \bibinfo {pages} {813} (\bibinfo {year} {2023})},\ \Eprint
  {https://arxiv.org/abs/2207.05212} {arXiv:2207.05212 [nucl-ex]} \BibitemShut
  {NoStop}%
\bibitem [{\citenamefont {Liu}\ and\ \citenamefont
  {Zahed}(2024)}]{Liu:2024yqa}%
  \BibitemOpen
  \bibfield  {author} {\bibinfo {author} {\bibfnamefont {W.-Y.}\ \bibnamefont
  {Liu}}\ and\ \bibinfo {author} {\bibfnamefont {I.}~\bibnamefont {Zahed}},\
  }\href {https://doi.org/10.1103/PhysRevD.110.054025} {\bibfield  {journal}
  {\bibinfo  {journal} {Phys. Rev. D}\ }\textbf {\bibinfo {volume} {110}},\
  \bibinfo {pages} {054025} (\bibinfo {year} {2024})},\ \Eprint
  {https://arxiv.org/abs/2404.03875} {arXiv:2404.03875 [hep-ph]} \BibitemShut
  {NoStop}%
\bibitem [{\citenamefont {Hechenberger}\ \emph {et~al.}(2024)\citenamefont
  {Hechenberger}, \citenamefont {Mamo},\ and\ \citenamefont
  {Zahed}}]{Hechenberger:2024abg}%
  \BibitemOpen
  \bibfield  {author} {\bibinfo {author} {\bibfnamefont {F.}~\bibnamefont
  {Hechenberger}}, \bibinfo {author} {\bibfnamefont {K.~A.}\ \bibnamefont
  {Mamo}},\ and\ \bibinfo {author} {\bibfnamefont {I.}~\bibnamefont {Zahed}},\
  }\href {https://doi.org/10.1103/PhysRevD.109.074013} {\bibfield  {journal}
  {\bibinfo  {journal} {Phys. Rev. D}\ }\textbf {\bibinfo {volume} {109}},\
  \bibinfo {pages} {074013} (\bibinfo {year} {2024})},\ \Eprint
  {https://arxiv.org/abs/2401.12162} {arXiv:2401.12162 [hep-ph]} \BibitemShut
  {NoStop}%
\bibitem [{\citenamefont {Guo}\ \emph {et~al.}(2021)\citenamefont {Guo},
  \citenamefont {Ji},\ and\ \citenamefont {Liu}}]{Guo:2021ibg}%
  \BibitemOpen
  \bibfield  {author} {\bibinfo {author} {\bibfnamefont {Y.}~\bibnamefont
  {Guo}}, \bibinfo {author} {\bibfnamefont {X.}~\bibnamefont {Ji}},\ and\
  \bibinfo {author} {\bibfnamefont {Y.}~\bibnamefont {Liu}},\ }\href
  {https://doi.org/10.1103/PhysRevD.103.096010} {\bibfield  {journal} {\bibinfo
   {journal} {Phys. Rev. D}\ }\textbf {\bibinfo {volume} {103}},\ \bibinfo
  {pages} {096010} (\bibinfo {year} {2021})},\ \Eprint
  {https://arxiv.org/abs/2103.11506} {arXiv:2103.11506 [hep-ph]} \BibitemShut
  {NoStop}%
\bibitem [{\citenamefont {Hatta}\ and\ \citenamefont
  {Strikman}(2021)}]{Hatta:2021can}%
  \BibitemOpen
  \bibfield  {author} {\bibinfo {author} {\bibfnamefont {Y.}~\bibnamefont
  {Hatta}}\ and\ \bibinfo {author} {\bibfnamefont {M.}~\bibnamefont
  {Strikman}},\ }\href {https://doi.org/10.1016/j.physletb.2021.136295}
  {\bibfield  {journal} {\bibinfo  {journal} {Phys. Lett. B}\ }\textbf
  {\bibinfo {volume} {817}},\ \bibinfo {pages} {136295} (\bibinfo {year}
  {2021})},\ \Eprint {https://arxiv.org/abs/2102.12631} {arXiv:2102.12631
  [hep-ph]} \BibitemShut {NoStop}%
\bibitem [{\citenamefont {Pefkou}\ \emph {et~al.}(2022)\citenamefont {Pefkou},
  \citenamefont {Hackett},\ and\ \citenamefont {Shanahan}}]{Pefkou:2021fni}%
  \BibitemOpen
  \bibfield  {author} {\bibinfo {author} {\bibfnamefont {D.~A.}\ \bibnamefont
  {Pefkou}}, \bibinfo {author} {\bibfnamefont {D.~C.}\ \bibnamefont
  {Hackett}},\ and\ \bibinfo {author} {\bibfnamefont {P.~E.}\ \bibnamefont
  {Shanahan}},\ }\href {https://doi.org/10.1103/PhysRevD.105.054509} {\bibfield
   {journal} {\bibinfo  {journal} {Phys. Rev. D}\ }\textbf {\bibinfo {volume}
  {105}},\ \bibinfo {pages} {054509} (\bibinfo {year} {2022})},\ \Eprint
  {https://arxiv.org/abs/2107.10368} {arXiv:2107.10368 [hep-lat]} \BibitemShut
  {NoStop}%
\bibitem [{\citenamefont {Hackett}\ \emph {et~al.}(2024)\citenamefont
  {Hackett}, \citenamefont {Pefkou},\ and\ \citenamefont
  {Shanahan}}]{Hackett:2023rif}%
  \BibitemOpen
  \bibfield  {author} {\bibinfo {author} {\bibfnamefont {D.~C.}\ \bibnamefont
  {Hackett}}, \bibinfo {author} {\bibfnamefont {D.~A.}\ \bibnamefont
  {Pefkou}},\ and\ \bibinfo {author} {\bibfnamefont {P.~E.}\ \bibnamefont
  {Shanahan}},\ }\href {https://doi.org/10.1103/PhysRevLett.132.251904}
  {\bibfield  {journal} {\bibinfo  {journal} {Phys. Rev. Lett.}\ }\textbf
  {\bibinfo {volume} {132}},\ \bibinfo {pages} {251904} (\bibinfo {year}
  {2024})},\ \Eprint {https://arxiv.org/abs/2310.08484} {arXiv:2310.08484
  [hep-lat]} \BibitemShut {NoStop}%
\bibitem [{\citenamefont {Hu}\ \emph {et~al.}(2024)\citenamefont {Hu},
  \citenamefont {Cao}, \citenamefont {Xu}, \citenamefont {Li}, \citenamefont
  {Zhao},\ and\ \citenamefont {Vary}}]{Hu:2024edc}%
  \BibitemOpen
  \bibfield  {author} {\bibinfo {author} {\bibfnamefont {T.}~\bibnamefont
  {Hu}}, \bibinfo {author} {\bibfnamefont {X.}~\bibnamefont {Cao}}, \bibinfo
  {author} {\bibfnamefont {S.}~\bibnamefont {Xu}}, \bibinfo {author}
  {\bibfnamefont {Y.}~\bibnamefont {Li}}, \bibinfo {author} {\bibfnamefont
  {X.}~\bibnamefont {Zhao}},\ and\ \bibinfo {author} {\bibfnamefont {J.~P.}\
  \bibnamefont {Vary}}\ }\href {https://doi.org/10.48550/arXiv.2408.09689}
  {10.48550/arXiv.2408.09689} (\bibinfo {year} {2024}),\ \Eprint
  {https://arxiv.org/abs/2408.09689} {arXiv:2408.09689 [hep-ph]} \BibitemShut
  {NoStop}%
\bibitem [{\citenamefont {Shanahan}\ and\ \citenamefont
  {Detmold}(2019)}]{Shanahan:2018nnv}%
  \BibitemOpen
  \bibfield  {author} {\bibinfo {author} {\bibfnamefont {P.~E.}\ \bibnamefont
  {Shanahan}}\ and\ \bibinfo {author} {\bibfnamefont {W.}~\bibnamefont
  {Detmold}},\ }\href {https://doi.org/10.1103/PhysRevLett.122.072003}
  {\bibfield  {journal} {\bibinfo  {journal} {Phys. Rev. Lett.}\ }\textbf
  {\bibinfo {volume} {122}},\ \bibinfo {pages} {072003} (\bibinfo {year}
  {2019})},\ \Eprint {https://arxiv.org/abs/1810.07589} {arXiv:1810.07589
  [nucl-th]} \BibitemShut {NoStop}%
\bibitem [{\citenamefont {Alexandrou}\ \emph {et~al.}(2020)\citenamefont
  {Alexandrou}, \citenamefont {Bacchio}, \citenamefont {Constantinou},
  \citenamefont {Finkenrath}, \citenamefont {Hadjiyiannakou}, \citenamefont
  {Jansen}, \citenamefont {Koutsou}, \citenamefont {Panagopoulos},\ and\
  \citenamefont {Spanoudes}}]{Alexandrou:2020sml}%
  \BibitemOpen
  \bibfield  {author} {\bibinfo {author} {\bibfnamefont {C.}~\bibnamefont
  {Alexandrou}}, \bibinfo {author} {\bibfnamefont {S.}~\bibnamefont {Bacchio}},
  \bibinfo {author} {\bibfnamefont {M.}~\bibnamefont {Constantinou}}, \bibinfo
  {author} {\bibfnamefont {J.}~\bibnamefont {Finkenrath}}, \bibinfo {author}
  {\bibfnamefont {K.}~\bibnamefont {Hadjiyiannakou}}, \bibinfo {author}
  {\bibfnamefont {K.}~\bibnamefont {Jansen}}, \bibinfo {author} {\bibfnamefont
  {G.}~\bibnamefont {Koutsou}}, \bibinfo {author} {\bibfnamefont
  {H.}~\bibnamefont {Panagopoulos}},\ and\ \bibinfo {author} {\bibfnamefont
  {G.}~\bibnamefont {Spanoudes}},\ }\href
  {https://doi.org/10.1103/PhysRevD.101.094513} {\bibfield  {journal} {\bibinfo
   {journal} {Phys. Rev. D}\ }\textbf {\bibinfo {volume} {101}},\ \bibinfo
  {pages} {094513} (\bibinfo {year} {2020})},\ \Eprint
  {https://arxiv.org/abs/2003.08486} {arXiv:2003.08486 [hep-lat]} \BibitemShut
  {NoStop}%
\bibitem [{\citenamefont {Mamo}\ and\ \citenamefont
  {Zahed}(2021)}]{Mamo:2021krl}%
  \BibitemOpen
  \bibfield  {author} {\bibinfo {author} {\bibfnamefont {K.~A.}\ \bibnamefont
  {Mamo}}\ and\ \bibinfo {author} {\bibfnamefont {I.}~\bibnamefont {Zahed}},\
  }\href {https://doi.org/10.1103/PhysRevD.103.094010} {\bibfield  {journal}
  {\bibinfo  {journal} {Phys. Rev. D}\ }\textbf {\bibinfo {volume} {103}},\
  \bibinfo {pages} {094010} (\bibinfo {year} {2021})},\ \Eprint
  {https://arxiv.org/abs/2103.03186} {arXiv:2103.03186 [hep-ph]} \BibitemShut
  {NoStop}%
\bibitem [{\citenamefont {Yao}\ \emph {et~al.}(2024)\citenamefont {Yao},
  \citenamefont {Xu}, \citenamefont {Binosi}, \citenamefont {Cui},
  \citenamefont {Ding}, \citenamefont {Raya}, \citenamefont {Roberts},
  \citenamefont {Rodr\'\i{}guez-Quintero},\ and\ \citenamefont
  {Schmidt}}]{Yao:2024ixu}%
  \BibitemOpen
  \bibfield  {author} {\bibinfo {author} {\bibfnamefont {Z.~Q.}\ \bibnamefont
  {Yao}}, \bibinfo {author} {\bibfnamefont {Y.~Z.}\ \bibnamefont {Xu}},
  \bibinfo {author} {\bibfnamefont {D.}~\bibnamefont {Binosi}}, \bibinfo
  {author} {\bibfnamefont {Z.~F.}\ \bibnamefont {Cui}}, \bibinfo {author}
  {\bibfnamefont {M.}~\bibnamefont {Ding}}, \bibinfo {author} {\bibfnamefont
  {K.}~\bibnamefont {Raya}}, \bibinfo {author} {\bibfnamefont {C.~D.}\
  \bibnamefont {Roberts}}, \bibinfo {author} {\bibfnamefont {J.}~\bibnamefont
  {Rodr\'\i{}guez-Quintero}},\ and\ \bibinfo {author} {\bibfnamefont {S.~M.}\
  \bibnamefont {Schmidt}},\ }\href@noop {} {\  (\bibinfo {year} {2024})},\
  \Eprint {https://arxiv.org/abs/2409.15547} {arXiv:2409.15547 [hep-ph]}
  \BibitemShut {NoStop}%
\bibitem [{\citenamefont {Vary}\ \emph {et~al.}(2010)\citenamefont {Vary},
  \citenamefont {Honkanen}, \citenamefont {Li}, \citenamefont {Maris},
  \citenamefont {Brodsky}, \citenamefont {Harindranath}, \citenamefont
  {de~Teramond}, \citenamefont {Sternberg}, \citenamefont {Ng},\ and\
  \citenamefont {Yang}}]{Vary:2009gt}%
  \BibitemOpen
  \bibfield  {author} {\bibinfo {author} {\bibfnamefont {J.~P.}\ \bibnamefont
  {Vary}}, \bibinfo {author} {\bibfnamefont {H.}~\bibnamefont {Honkanen}},
  \bibinfo {author} {\bibfnamefont {J.}~\bibnamefont {Li}}, \bibinfo {author}
  {\bibfnamefont {P.}~\bibnamefont {Maris}}, \bibinfo {author} {\bibfnamefont
  {S.~J.}\ \bibnamefont {Brodsky}}, \bibinfo {author} {\bibfnamefont
  {A.}~\bibnamefont {Harindranath}}, \bibinfo {author} {\bibfnamefont {G.~F.}\
  \bibnamefont {de~Teramond}}, \bibinfo {author} {\bibfnamefont
  {P.}~\bibnamefont {Sternberg}}, \bibinfo {author} {\bibfnamefont {E.~G.}\
  \bibnamefont {Ng}},\ and\ \bibinfo {author} {\bibfnamefont {C.}~\bibnamefont
  {Yang}},\ }\href {https://doi.org/10.1103/PhysRevC.81.035205} {\bibfield
  {journal} {\bibinfo  {journal} {Phys. Rev. C}\ }\textbf {\bibinfo {volume}
  {81}},\ \bibinfo {pages} {035205} (\bibinfo {year} {2010})},\ \Eprint
  {https://arxiv.org/abs/0905.1411} {arXiv:0905.1411 [nucl-th]} \BibitemShut
  {NoStop}%
\bibitem [{\citenamefont {Mondal}\ \emph {et~al.}(2020)\citenamefont {Mondal},
  \citenamefont {Xu}, \citenamefont {Lan}, \citenamefont {Zhao}, \citenamefont
  {Li}, \citenamefont {Chakrabarti},\ and\ \citenamefont
  {Vary}}]{Mondal:2019jdg}%
  \BibitemOpen
  \bibfield  {author} {\bibinfo {author} {\bibfnamefont {C.}~\bibnamefont
  {Mondal}}, \bibinfo {author} {\bibfnamefont {S.}~\bibnamefont {Xu}}, \bibinfo
  {author} {\bibfnamefont {J.}~\bibnamefont {Lan}}, \bibinfo {author}
  {\bibfnamefont {X.}~\bibnamefont {Zhao}}, \bibinfo {author} {\bibfnamefont
  {Y.}~\bibnamefont {Li}}, \bibinfo {author} {\bibfnamefont {D.}~\bibnamefont
  {Chakrabarti}},\ and\ \bibinfo {author} {\bibfnamefont {J.~P.}\ \bibnamefont
  {Vary}} (\bibinfo {collaboration} {BLFQ}),\ }\href
  {https://doi.org/10.1103/PhysRevD.102.016008} {\bibfield  {journal} {\bibinfo
   {journal} {Phys. Rev. D}\ }\textbf {\bibinfo {volume} {102}},\ \bibinfo
  {pages} {016008} (\bibinfo {year} {2020})},\ \Eprint
  {https://arxiv.org/abs/1911.10913} {arXiv:1911.10913 [hep-ph]} \BibitemShut
  {NoStop}%
\bibitem [{\citenamefont {Xu}\ \emph {et~al.}(2021)\citenamefont {Xu},
  \citenamefont {Mondal}, \citenamefont {Lan}, \citenamefont {Zhao},
  \citenamefont {Li},\ and\ \citenamefont {Vary}}]{Xu:2021wwj}%
  \BibitemOpen
  \bibfield  {author} {\bibinfo {author} {\bibfnamefont {S.}~\bibnamefont
  {Xu}}, \bibinfo {author} {\bibfnamefont {C.}~\bibnamefont {Mondal}}, \bibinfo
  {author} {\bibfnamefont {J.}~\bibnamefont {Lan}}, \bibinfo {author}
  {\bibfnamefont {X.}~\bibnamefont {Zhao}}, \bibinfo {author} {\bibfnamefont
  {Y.}~\bibnamefont {Li}},\ and\ \bibinfo {author} {\bibfnamefont {J.~P.}\
  \bibnamefont {Vary}} (\bibinfo {collaboration} {BLFQ}),\ }\href
  {https://doi.org/10.1103/PhysRevD.104.094036} {\bibfield  {journal} {\bibinfo
   {journal} {Phys. Rev. D}\ }\textbf {\bibinfo {volume} {104}},\ \bibinfo
  {pages} {094036} (\bibinfo {year} {2021})},\ \Eprint
  {https://arxiv.org/abs/2108.03909} {arXiv:2108.03909 [hep-ph]} \BibitemShut
  {NoStop}%
\bibitem [{\citenamefont {Xu}\ \emph {et~al.}(2023{\natexlab{a}})\citenamefont
  {Xu}, \citenamefont {Mondal}, \citenamefont {Zhao}, \citenamefont {Li},\ and\
  \citenamefont {Vary}}]{Xu:2022yxb}%
  \BibitemOpen
  \bibfield  {author} {\bibinfo {author} {\bibfnamefont {S.}~\bibnamefont
  {Xu}}, \bibinfo {author} {\bibfnamefont {C.}~\bibnamefont {Mondal}}, \bibinfo
  {author} {\bibfnamefont {X.}~\bibnamefont {Zhao}}, \bibinfo {author}
  {\bibfnamefont {Y.}~\bibnamefont {Li}},\ and\ \bibinfo {author}
  {\bibfnamefont {J.~P.}\ \bibnamefont {Vary}} (\bibinfo {collaboration}
  {BLFQ}),\ }\href {https://doi.org/10.1103/PhysRevD.108.094002} {\bibfield
  {journal} {\bibinfo  {journal} {Phys. Rev. D}\ }\textbf {\bibinfo {volume}
  {108}},\ \bibinfo {pages} {094002} (\bibinfo {year} {2023}{\natexlab{a}})},\
  \Eprint {https://arxiv.org/abs/2209.08584} {arXiv:2209.08584 [hep-ph]}
  \BibitemShut {NoStop}%
\bibitem [{\citenamefont {Brodsky}\ \emph {et~al.}(1998)\citenamefont
  {Brodsky}, \citenamefont {Pauli},\ and\ \citenamefont
  {Pinsky}}]{Brodsky:1997de}%
  \BibitemOpen
  \bibfield  {author} {\bibinfo {author} {\bibfnamefont {S.~J.}\ \bibnamefont
  {Brodsky}}, \bibinfo {author} {\bibfnamefont {H.-C.}\ \bibnamefont {Pauli}},\
  and\ \bibinfo {author} {\bibfnamefont {S.~S.}\ \bibnamefont {Pinsky}},\
  }\href {https://doi.org/10.1016/S0370-1573(97)00089-6} {\bibfield  {journal}
  {\bibinfo  {journal} {Phys. Rept.}\ }\textbf {\bibinfo {volume} {301}},\
  \bibinfo {pages} {299} (\bibinfo {year} {1998})},\ \Eprint
  {https://arxiv.org/abs/hep-ph/9705477} {arXiv:hep-ph/9705477} \BibitemShut
  {NoStop}%
\bibitem [{\citenamefont {Li}\ \emph {et~al.}(2016)\citenamefont {Li},
  \citenamefont {Maris}, \citenamefont {Zhao},\ and\ \citenamefont
  {Vary}}]{Li:2015zda}%
  \BibitemOpen
  \bibfield  {author} {\bibinfo {author} {\bibfnamefont {Y.}~\bibnamefont
  {Li}}, \bibinfo {author} {\bibfnamefont {P.}~\bibnamefont {Maris}}, \bibinfo
  {author} {\bibfnamefont {X.}~\bibnamefont {Zhao}},\ and\ \bibinfo {author}
  {\bibfnamefont {J.~P.}\ \bibnamefont {Vary}},\ }\href
  {https://doi.org/10.1016/j.physletb.2016.04.065} {\bibfield  {journal}
  {\bibinfo  {journal} {Phys. Lett. B}\ }\textbf {\bibinfo {volume} {758}},\
  \bibinfo {pages} {118} (\bibinfo {year} {2016})},\ \Eprint
  {https://arxiv.org/abs/1509.07212} {arXiv:1509.07212 [hep-ph]} \BibitemShut
  {NoStop}%
\bibitem [{\citenamefont {Xu}\ \emph {et~al.}(2023{\natexlab{b}})\citenamefont
  {Xu}, \citenamefont {Mondal}, \citenamefont {Zhao}, \citenamefont {Li},\ and\
  \citenamefont {Vary}}]{Xu:2023nqv}%
  \BibitemOpen
  \bibfield  {author} {\bibinfo {author} {\bibfnamefont {S.}~\bibnamefont
  {Xu}}, \bibinfo {author} {\bibfnamefont {C.}~\bibnamefont {Mondal}}, \bibinfo
  {author} {\bibfnamefont {X.}~\bibnamefont {Zhao}}, \bibinfo {author}
  {\bibfnamefont {Y.}~\bibnamefont {Li}},\ and\ \bibinfo {author}
  {\bibfnamefont {J.~P.}\ \bibnamefont {Vary}} (\bibinfo {collaboration}
  {BLFQ}),\ }\href {https://doi.org/10.1103/PhysRevD.108.094002} {\bibfield
  {journal} {\bibinfo  {journal} {Phys. Rev. D}\ }\textbf {\bibinfo {volume}
  {108}},\ \bibinfo {pages} {094002} (\bibinfo {year}
  {2023}{\natexlab{b}})}\BibitemShut {NoStop}%
\bibitem [{\citenamefont {Lan}\ \emph {et~al.}(2022)\citenamefont {Lan},
  \citenamefont {Fu}, \citenamefont {Mondal}, \citenamefont {Zhao},\ and\
  \citenamefont {Vary}}]{Lan:2021wok}%
  \BibitemOpen
  \bibfield  {author} {\bibinfo {author} {\bibfnamefont {J.}~\bibnamefont
  {Lan}}, \bibinfo {author} {\bibfnamefont {K.}~\bibnamefont {Fu}}, \bibinfo
  {author} {\bibfnamefont {C.}~\bibnamefont {Mondal}}, \bibinfo {author}
  {\bibfnamefont {X.}~\bibnamefont {Zhao}},\ and\ \bibinfo {author}
  {\bibfnamefont {J.~P.}\ \bibnamefont {Vary}} (\bibinfo {collaboration}
  {BLFQ}),\ }\href {https://doi.org/10.1016/j.physletb.2022.136890} {\bibfield
  {journal} {\bibinfo  {journal} {Phys. Lett. B}\ }\textbf {\bibinfo {volume}
  {825}},\ \bibinfo {pages} {136890} (\bibinfo {year} {2022})},\ \Eprint
  {https://arxiv.org/abs/2106.04954} {arXiv:2106.04954 [hep-ph]} \BibitemShut
  {NoStop}%
\bibitem [{\citenamefont {Cornwall}(1982)}]{PhysRevD.26.1453}%
  \BibitemOpen
  \bibfield  {author} {\bibinfo {author} {\bibfnamefont {J.~M.}\ \bibnamefont
  {Cornwall}},\ }\href {https://doi.org/10.1103/PhysRevD.26.1453} {\bibfield
  {journal} {\bibinfo  {journal} {Phys. Rev. D}\ }\textbf {\bibinfo {volume}
  {26}},\ \bibinfo {pages} {1453} (\bibinfo {year} {1982})}\BibitemShut
  {NoStop}%
\bibitem [{\citenamefont {Perry}\ \emph {et~al.}(1990)\citenamefont {Perry},
  \citenamefont {Harindranath},\ and\ \citenamefont {Wilson}}]{Perry:1990mz}%
  \BibitemOpen
  \bibfield  {author} {\bibinfo {author} {\bibfnamefont {R.~J.}\ \bibnamefont
  {Perry}}, \bibinfo {author} {\bibfnamefont {A.}~\bibnamefont
  {Harindranath}},\ and\ \bibinfo {author} {\bibfnamefont {K.~G.}\ \bibnamefont
  {Wilson}},\ }\href {https://doi.org/10.1103/PhysRevLett.65.2959} {\bibfield
  {journal} {\bibinfo  {journal} {Phys. Rev. Lett.}\ }\textbf {\bibinfo
  {volume} {65}},\ \bibinfo {pages} {2959} (\bibinfo {year}
  {1990})}\BibitemShut {NoStop}%
\bibitem [{\citenamefont {Karmanov}\ \emph {et~al.}(2008)\citenamefont
  {Karmanov}, \citenamefont {Mathiot},\ and\ \citenamefont
  {Smirnov}}]{Karmanov:2008br}%
  \BibitemOpen
  \bibfield  {author} {\bibinfo {author} {\bibfnamefont {V.~A.}\ \bibnamefont
  {Karmanov}}, \bibinfo {author} {\bibfnamefont {J.~F.}\ \bibnamefont
  {Mathiot}},\ and\ \bibinfo {author} {\bibfnamefont {A.~V.}\ \bibnamefont
  {Smirnov}},\ }\href {https://doi.org/10.1103/PhysRevD.77.085028} {\bibfield
  {journal} {\bibinfo  {journal} {Phys. Rev. D}\ }\textbf {\bibinfo {volume}
  {77}},\ \bibinfo {pages} {085028} (\bibinfo {year} {2008})},\ \Eprint
  {https://arxiv.org/abs/0801.4507} {arXiv:0801.4507 [hep-th]} \BibitemShut
  {NoStop}%
\bibitem [{\citenamefont {Burkardt}(1998)}]{Burkardt:1998dd}%
  \BibitemOpen
  \bibfield  {author} {\bibinfo {author} {\bibfnamefont {M.}~\bibnamefont
  {Burkardt}},\ }\href {https://doi.org/10.1103/PhysRevD.58.096015} {\bibfield
  {journal} {\bibinfo  {journal} {Phys. Rev. D}\ }\textbf {\bibinfo {volume}
  {58}},\ \bibinfo {pages} {096015} (\bibinfo {year} {1998})},\ \Eprint
  {https://arxiv.org/abs/hep-th/9805088} {arXiv:hep-th/9805088} \BibitemShut
  {NoStop}%
\bibitem [{\citenamefont {Zhao}\ \emph {et~al.}(2014)\citenamefont {Zhao},
  \citenamefont {Honkanen}, \citenamefont {Maris}, \citenamefont {Vary},\ and\
  \citenamefont {Brodsky}}]{Zhao:2014xaa}%
  \BibitemOpen
  \bibfield  {author} {\bibinfo {author} {\bibfnamefont {X.}~\bibnamefont
  {Zhao}}, \bibinfo {author} {\bibfnamefont {H.}~\bibnamefont {Honkanen}},
  \bibinfo {author} {\bibfnamefont {P.}~\bibnamefont {Maris}}, \bibinfo
  {author} {\bibfnamefont {J.~P.}\ \bibnamefont {Vary}},\ and\ \bibinfo
  {author} {\bibfnamefont {S.~J.}\ \bibnamefont {Brodsky}},\ }\href
  {https://doi.org/10.1016/j.physletb.2014.08.020} {\bibfield  {journal}
  {\bibinfo  {journal} {Phys. Lett. B}\ }\textbf {\bibinfo {volume} {737}},\
  \bibinfo {pages} {65} (\bibinfo {year} {2014})},\ \Eprint
  {https://arxiv.org/abs/1402.4195} {arXiv:1402.4195 [nucl-th]} \BibitemShut
  {NoStop}%
\bibitem [{\citenamefont {Bhattacharya}\ \emph {et~al.}(2023)\citenamefont
  {Bhattacharya}, \citenamefont {Cichy}, \citenamefont {Constantinou},
  \citenamefont {Gao}, \citenamefont {Metz}, \citenamefont {Miller},
  \citenamefont {Mukherjee}, \citenamefont {Petreczky}, \citenamefont
  {Steffens},\ and\ \citenamefont {Zhao}}]{Bhattacharya:2023ays}%
  \BibitemOpen
  \bibfield  {author} {\bibinfo {author} {\bibfnamefont {S.}~\bibnamefont
  {Bhattacharya}}, \bibinfo {author} {\bibfnamefont {K.}~\bibnamefont {Cichy}},
  \bibinfo {author} {\bibfnamefont {M.}~\bibnamefont {Constantinou}}, \bibinfo
  {author} {\bibfnamefont {X.}~\bibnamefont {Gao}}, \bibinfo {author}
  {\bibfnamefont {A.}~\bibnamefont {Metz}}, \bibinfo {author} {\bibfnamefont
  {J.}~\bibnamefont {Miller}}, \bibinfo {author} {\bibfnamefont
  {S.}~\bibnamefont {Mukherjee}}, \bibinfo {author} {\bibfnamefont
  {P.}~\bibnamefont {Petreczky}}, \bibinfo {author} {\bibfnamefont
  {F.}~\bibnamefont {Steffens}},\ and\ \bibinfo {author} {\bibfnamefont
  {Y.}~\bibnamefont {Zhao}},\ }\href
  {https://doi.org/10.1103/PhysRevD.108.014507} {\bibfield  {journal} {\bibinfo
   {journal} {Phys. Rev. D}\ }\textbf {\bibinfo {volume} {108}},\ \bibinfo
  {pages} {014507} (\bibinfo {year} {2023})},\ \Eprint
  {https://arxiv.org/abs/2305.11117} {arXiv:2305.11117 [hep-lat]} \BibitemShut
  {NoStop}%
\bibitem [{\citenamefont {Wang}\ \emph {et~al.}(2024)\citenamefont {Wang},
  \citenamefont {Zeng},\ and\ \citenamefont {Zhang}}]{Wang:2023fmx}%
  \BibitemOpen
  \bibfield  {author} {\bibinfo {author} {\bibfnamefont {X.-Y.}\ \bibnamefont
  {Wang}}, \bibinfo {author} {\bibfnamefont {F.}~\bibnamefont {Zeng}},\ and\
  \bibinfo {author} {\bibfnamefont {J.}~\bibnamefont {Zhang}},\ }\href
  {https://doi.org/10.1088/1674-1137/ad2a66} {\bibfield  {journal} {\bibinfo
  {journal} {Chin. Phys. C}\ }\textbf {\bibinfo {volume} {48}},\ \bibinfo
  {pages} {054102} (\bibinfo {year} {2024})},\ \Eprint
  {https://arxiv.org/abs/2308.04644} {arXiv:2308.04644 [hep-ph]} \BibitemShut
  {NoStop}%
\bibitem [{\citenamefont {Mamo}\ and\ \citenamefont
  {Zahed}(2024)}]{Mamo:2024jwp}%
  \BibitemOpen
  \bibfield  {author} {\bibinfo {author} {\bibfnamefont {K.~A.}\ \bibnamefont
  {Mamo}}\ and\ \bibinfo {author} {\bibfnamefont {I.}~\bibnamefont {Zahed}},\
  }\href {https://doi.org/10.1103/PhysRevLett.133.241901} {\bibfield  {journal}
  {\bibinfo  {journal} {Phys. Rev. Lett.}\ }\textbf {\bibinfo {volume} {133}},\
  \bibinfo {pages} {241901} (\bibinfo {year} {2024})},\ \Eprint
  {https://arxiv.org/abs/2411.04162} {arXiv:2411.04162 [hep-ph]} \BibitemShut
  {NoStop}%
\bibitem [{\citenamefont {Guo}\ \emph {et~al.}(2023)\citenamefont {Guo},
  \citenamefont {Ji}, \citenamefont {Liu},\ and\ \citenamefont
  {Yang}}]{Guo:2023pqw}%
  \BibitemOpen
  \bibfield  {author} {\bibinfo {author} {\bibfnamefont {Y.}~\bibnamefont
  {Guo}}, \bibinfo {author} {\bibfnamefont {X.}~\bibnamefont {Ji}}, \bibinfo
  {author} {\bibfnamefont {Y.}~\bibnamefont {Liu}},\ and\ \bibinfo {author}
  {\bibfnamefont {J.}~\bibnamefont {Yang}},\ }\href
  {https://doi.org/10.1103/PhysRevD.108.034003} {\bibfield  {journal} {\bibinfo
   {journal} {Phys. Rev. D}\ }\textbf {\bibinfo {volume} {108}},\ \bibinfo
  {pages} {034003} (\bibinfo {year} {2023})},\ \Eprint
  {https://arxiv.org/abs/2305.06992} {arXiv:2305.06992 [hep-ph]} \BibitemShut
  {NoStop}%
\bibitem [{\citenamefont {Kim}\ and\ \citenamefont {Kim}(2021)}]{Kim:2021jjf}%
  \BibitemOpen
  \bibfield  {author} {\bibinfo {author} {\bibfnamefont {J.-Y.}\ \bibnamefont
  {Kim}}\ and\ \bibinfo {author} {\bibfnamefont {H.-C.}\ \bibnamefont {Kim}},\
  }\href {https://doi.org/10.1103/PhysRevD.104.074019} {\bibfield  {journal}
  {\bibinfo  {journal} {Phys. Rev. D}\ }\textbf {\bibinfo {volume} {104}},\
  \bibinfo {pages} {074019} (\bibinfo {year} {2021})},\ \Eprint
  {https://arxiv.org/abs/2105.10279} {arXiv:2105.10279 [hep-ph]} \BibitemShut
  {NoStop}%
\bibitem [{\citenamefont {Navas}\ \emph {et~al.}(2024)\citenamefont {Navas}
  \emph {et~al.}}]{ParticleDataGroup:2024cfk}%
  \BibitemOpen
  \bibfield  {author} {\bibinfo {author} {\bibfnamefont {S.}~\bibnamefont
  {Navas}} \emph {et~al.} (\bibinfo {collaboration} {Particle Data Group}),\
  }\href {https://doi.org/10.1103/PhysRevD.110.030001} {\bibfield  {journal}
  {\bibinfo  {journal} {Phys. Rev. D}\ }\textbf {\bibinfo {volume} {110}},\
  \bibinfo {pages} {030001} (\bibinfo {year} {2024})}\BibitemShut {NoStop}%
\bibitem [{\citenamefont {Ji}(1995)}]{Ji:1994av}%
  \BibitemOpen
  \bibfield  {author} {\bibinfo {author} {\bibfnamefont {X.-D.}\ \bibnamefont
  {Ji}},\ }\href {https://doi.org/10.1103/PhysRevLett.74.1071} {\bibfield
  {journal} {\bibinfo  {journal} {Phys. Rev. Lett.}\ }\textbf {\bibinfo
  {volume} {74}},\ \bibinfo {pages} {1071} (\bibinfo {year} {1995})},\ \Eprint
  {https://arxiv.org/abs/hep-ph/9410274} {arXiv:hep-ph/9410274} \BibitemShut
  {NoStop}%
\bibitem [{\citenamefont {Lorc\'e}(2018)}]{Lorce:2017xzd}%
  \BibitemOpen
  \bibfield  {author} {\bibinfo {author} {\bibfnamefont {C.}~\bibnamefont
  {Lorc\'e}},\ }\href {https://doi.org/10.1140/epjc/s10052-018-5561-2}
  {\bibfield  {journal} {\bibinfo  {journal} {Eur. Phys. J. C}\ }\textbf
  {\bibinfo {volume} {78}},\ \bibinfo {pages} {120} (\bibinfo {year} {2018})},\
  \Eprint {https://arxiv.org/abs/1706.05853} {arXiv:1706.05853 [hep-ph]}
  \BibitemShut {NoStop}%
\bibitem [{\citenamefont {Yang}\ \emph {et~al.}(2018)\citenamefont {Yang},
  \citenamefont {Liang}, \citenamefont {Bi}, \citenamefont {Chen},
  \citenamefont {Draper}, \citenamefont {Liu},\ and\ \citenamefont
  {Liu}}]{Yang:2018nqn}%
  \BibitemOpen
  \bibfield  {author} {\bibinfo {author} {\bibfnamefont {Y.-B.}\ \bibnamefont
  {Yang}}, \bibinfo {author} {\bibfnamefont {J.}~\bibnamefont {Liang}},
  \bibinfo {author} {\bibfnamefont {Y.-J.}\ \bibnamefont {Bi}}, \bibinfo
  {author} {\bibfnamefont {Y.}~\bibnamefont {Chen}}, \bibinfo {author}
  {\bibfnamefont {T.}~\bibnamefont {Draper}}, \bibinfo {author} {\bibfnamefont
  {K.-F.}\ \bibnamefont {Liu}},\ and\ \bibinfo {author} {\bibfnamefont
  {Z.}~\bibnamefont {Liu}},\ }\href
  {https://doi.org/10.1103/PhysRevLett.121.212001} {\bibfield  {journal}
  {\bibinfo  {journal} {Phys. Rev. Lett.}\ }\textbf {\bibinfo {volume} {121}},\
  \bibinfo {pages} {212001} (\bibinfo {year} {2018})},\ \Eprint
  {https://arxiv.org/abs/1808.08677} {arXiv:1808.08677 [hep-lat]} \BibitemShut
  {NoStop}%
\end{thebibliography}%
\pagestyle{empty}
\begin{center}
	\hspace*{-11.0cm}
	\includegraphics[page=1,width=\paperwidth,height=\paperheight,keepaspectratio]{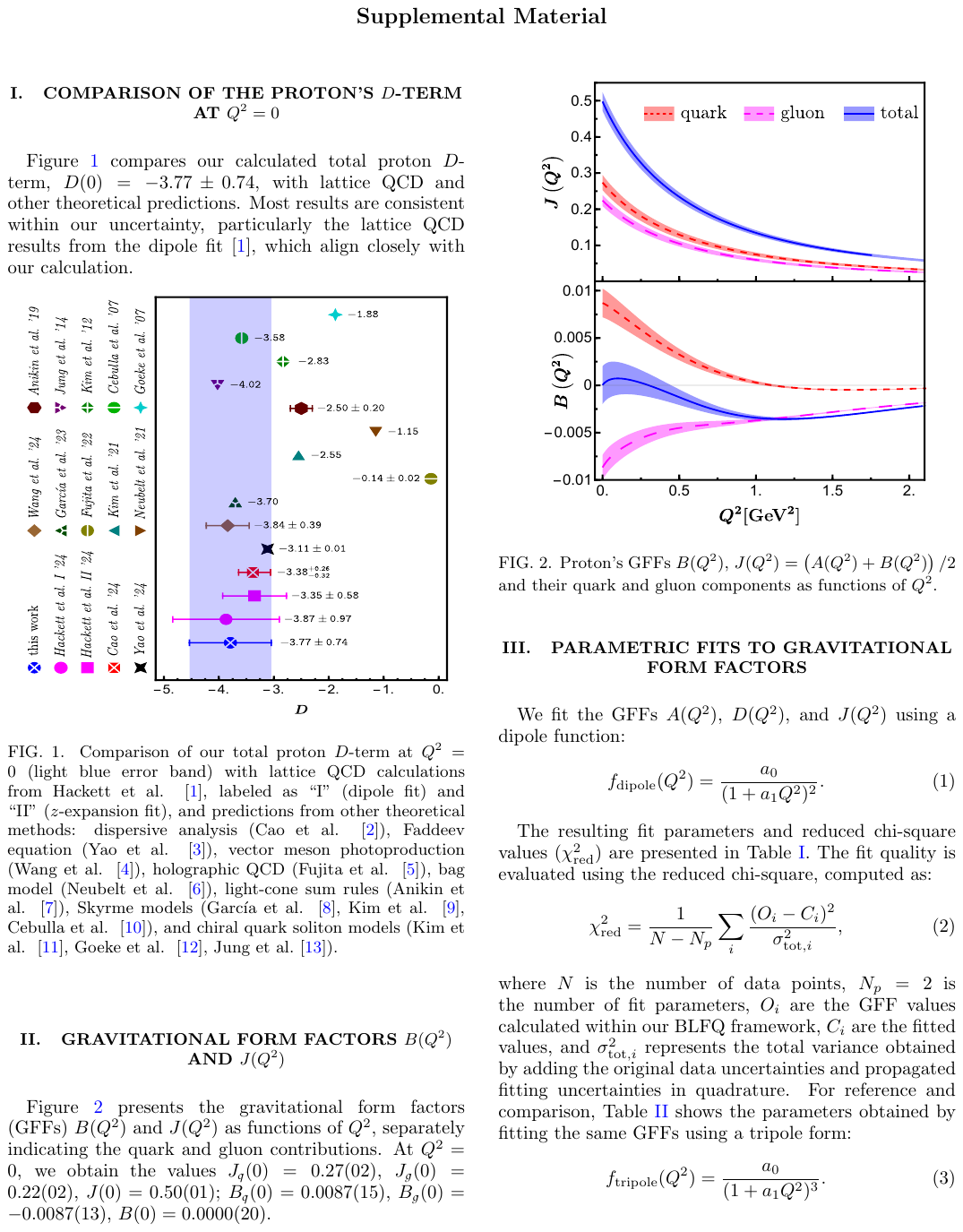}
		\hspace*{-11.0cm}
		\includegraphics[page=2,width=\paperwidth,height=\paperheight,keepaspectratio]{supplemental.pdf}
\end{center}
\end{document}